\documentclass[aps,onecolumn,notitlepage,showpacs,superscriptaddress]{revtex4-1}

\usepackage{graphicx}
\usepackage{bm}
\usepackage{verbatim}
\usepackage[section]{placeins}
\usepackage[normalsize,bf]{subfigure}
\usepackage{amsmath}

\begin{document}

\title{\large{Intrinsic noise and two-dimensional maps: quasi-cycles, quasiperiodicity and chaos}}

\author{C\'{e}sar Parra-Rojas}
\email{cesar.parrarojas@postgrad.manchester.ac.uk}
\affiliation{Theoretical Physics Division, School of Physics and Astronomy, 
The University of Manchester, Manchester M13 9PL, UK}
\author{Joseph~D. Challenger}
\email{jchallenger@unifi.it}
\author{Duccio Fanelli}
\email{duccio.fanelli@unifi.it}
\affiliation{Dipartimento di Fisica e Astronomia, Universit\`{a} degli Studi di Firenze and INFN, via Sansone 1, IT 50019 Sesto Fiorentino, Italy}
\author{Alan~J. McKane}
\email{alan.mckane@manchester.ac.uk}
\affiliation{Theoretical Physics Division, School of Physics and Astronomy, 
The University of Manchester, Manchester M13 9PL, UK}

\begin{abstract}
We develop a formalism to describe the discrete-time dynamics of systems containing an arbitrary number of interacting species. The individual-based model, which forms our starting point, is described by a Markov chain, which in the limit of large system sizes is shown to be very well-approximated by a Fokker-Planck-like equation, or equivalently by a set of stochastic difference equations. This formalism is applied to the specific case of two species: one predator species and its prey species. Quasi-cycles --- stochastic cycles sustained and amplified by the demographic noise --- previously found in continuous-time predator-prey models are shown to exist, and their behavior predicted from a linear noise analysis is shown to be in very good agreement with simulations. The effects of the noise on other attractors in the corresponding deterministic map, such as periodic cycles, quasiperiodicity and chaos, are also investigated.
\end{abstract}


\maketitle

\section{Introduction}
\label{sec:intro}

In a wide variety of disciplines, the mathematical modeling of population dynamics is a much-used tool. This can be useful to measure, for instance, the spread of a disease inside a community, or the abundances of biochemical molecules in a metabolic process. When setting up such a model, one is often faced with the choice of whether the model should depict the individual components present in the system (be they molecules, animals, or other entities), or simply describe the population by a macroscopic concentration. A popular choice is to take a mesoscopic approach, where instead of tracking each individual one is content to describe the fraction of individuals of each `species' present in the system \cite{black_12,goutsias_13}. The stochasticity, present in the interactions between individual elements in the system is, however, retained. Typically, the time-evolution of the population is taken to be a Markov process, that is, memoryless.

In the formalism described above, time is usually treated as a continuous variable, and the system can be described by a master equation \cite{kampen_07}. However, there are occasions where it is preferable to treat time as a discrete variable. This is often done in ecology, when studying the time-evolution of species which have non-overlapping generations, or in cases where field data is only collected at fixed time intervals. Therefore, to describe the evolution of the system, discrete time is more suitable. Nevertheless, the mesoscopic, or `individual-based' approach to modeling, carried out so frequently in continuous-time systems, has not been developed for discrete-time systems. Instead, such systems are typically described by deterministic difference equations, often referred to as `maps'\cite{strogatz_94}. Or, if stochasticity is found in the models, it is put in `by hand' through an additive noise term (see e.g. Refs.~\cite{mayer_81,crutchfield_82,gao_99}), rather than stemming from the underlying interactions between the individuals.

Our recent work has attempted to fill this gap, by developing a mesoscopic description of discrete-time population models \cite{challenger_13,challenger_14}, but for systems with only one variable. The corresponding deterministic description of the system is recovered in the thermodynamic limit, that is, in the limit of infinite population size. In this article, we present the extension of the theory to an arbitrary number of variables, although the application of the theory will be to a system with two variables (in what follows, we shall sometimes refer to systems with $d$ variables as $d$-dimensional). This extension is desirable for a number of reasons. Two-dimensional (2D) discrete-time models are particularly prevalent in ecology, where they are used to model interactions between two species. These interactions could take the form of competition between species, predator-prey or host-parasitoid dynamics (see \emph{e.g.} Refs.~\cite{hassell_76,may_76,sole_92c,beddington_75,beddington_76,lauwerier_86a,lauwerier_86b,may_87,sole_92a,sole_92b,neubert_92}). Furthermore, 2D maps display a greater range of different behavior than 1D maps, such as quasiperiodic attractors. Similar comments apply \textit{a fortiori} to higher-dimensional systems.

From a theoretical viewpoint, then, it is interesting to examine the effects of intrinsic noise in these systems. In continuous-time systems with dimension greater than one, it is known that the inclusion of stochasticity in the model can lead to sustained oscillations which are not predicted by the deterministic equations \cite{mckane_05}. The occurrence of these oscillations requires that the deterministic model approaches a stable fixed point \emph{via} damped oscillations, behavior which can be anticipated by calculating the eigenvalues of the map's Jacobian, evaluated at the fixed point. So, it is interesting to ask whether such oscillations, often called `quasi-cycles', appear in discrete-time systems.

Discrete-time systems have been frequently used to study chaotic behavior in a simple setting. In general, two main routes to chaos are observed in two-dimensional maps, one being a series of period-doubling bifurcations from a fixed point---already observed in 1D and generally present in competition models~\cite{hassell_76,may_76,sole_92c}; the other corresponds to a Hopf bifurcation leading to high-period or quasiperiodic trajectories, which then further bifurcate into a sequence of periodic states---periodic windows in the quasiperiodic region---with chaos emerging from one of them \emph{via} period-doubling or intermittency~\cite{kaneko_86}. The latter is the case in a wide range of predator-prey and host-parasite models~\cite{beddington_75,beddington_76,lauwerier_86a,lauwerier_86b,may_87,sole_92a,sole_92b,neubert_92}. In the literature, many papers have been devoted to the effects of noise in chaotic systems, including the observation that noise can induce chaos. For a parameter choice where the deterministic map is non-chaotic, the addition of noise can produce time series which have the hallmarks of chaos. The idea of noise-induced chaos has been much studied in the literature for the case of additive noise (see \emph{e.g.} Refs.~\cite{crutchfield_82,gao_99}). In a recent article~\cite{challenger_14} we investigated the ability of intrinsic noise to cause such a transition. Here we will do this for the case of 2D maps, and ascertain in which regions of parameter space this transition could be seen. An essential point here is that the noise is not additive: it has a multiplicative structure, the form of which can only be found by starting from an underlying microscopic model.

The article will be organized as follows. Section~\ref{sec:theory} introduces the Markov chain theory, which describes the microscopic process, before presenting the methodology that yields the mesoscopic description. This takes the form of a generalization of the Fokker-Planck equation, which can equivalently be formulated as a stochastic difference equation. In this section the formalism is presented for an arbitrary number of dimensions, and some of the technical details have been placed in the appendices.  We then show how a linearization of the stochastic difference equation can be used to obtain approximate, analytical predictions of the fluctuations around various attractors, such as a fixed point or $n$-cycle. Finally in Section~\ref{sec:theory} we derive a theoretical description for the power spectra of noise-induced oscillations, or `quasi-cycles', which are not predicted by the deterministic theory. Section~\ref{sec:lv} shows some of the results that can be obtained for the case of a map with a Lotka-Volterra-type dynamics. Here we quantify the fluctuations around a stable fixed point, as well as reporting the observation of quasi-cycles. In this section we also move beyond the fixed point, to look at the effects of intrinsic noise around periodic and quasiperiodic attractors. Section~\ref{sec:chaos} examines the ability of noise to induce chaotic behavior for parameter choices where the deterministic map is non-chaotic. Section~\ref{sec:disc} summarizes the article's findings, and suggests avenues for further work.

\section{The Mesoscopic Derivation}
\label{sec:theory}

In previous articles \cite{challenger_13,challenger_14} we developed a mesoscopic formulation of a microscopic Markov chain model. A 1D Markov chain is typically written as $P_{n,t+1}=\sum_{m}\,Q_{n m} P_{m,t}$, where $P_t$ is a vector of probabilities, each entry giving the probability for the system to be found in a particular state. In applications that we consider, the state of the system is simply the number of individuals in the system at a particular time, and is denoted by the integer $n$, where $n=0,1,\hdots,N$, $N$ being the maximum population that can be supported. Between two discrete times, $t$ and $t+1$, the probability vector is updated \emph{via} an application of the matrix $\bm{Q}$, which is known as the `transition matrix'. This matrix, along with an initial condition, $P_{t=0}$, defines the process. In our previous work, $\bm{Q}$ was given the following, binomial structure:
\begin{equation}
Q_{nm} = {N \choose n} p^{n} \left( 1 - p \right)^{N-n},
\label{binomial}
\end{equation}
where $p$ is a function of $m/N$, and where $m$ is the initial number of individuals present. In \cite{challenger_13} we showed that the $N \to \infty$ limit of the process has the form of a deterministic 1D map, governed by the function $p$. The subsequent article \cite{challenger_14} contained details of the derivation leading to a mesoscopic description for the system. This description is valid for finite $N$, where $N$ is not too small. In these two articles, we used the logistic map to motivate the form of $p$, but the method can be applied more generally. Here we will describe how this method can be extended to settings with $d$ distinct types of individuals, and discuss an application to a specific model containing two species. 

Suppose then that a system consists of individuals of $d$ distinct species, so that the state of the system is now $\bm{n}=(n_1,\ldots,n_d)$, where $n_i$ is the number of individuals of species $i$. Thus, we will write $P_{\bm{n},t}=P_{n_1n_2\hdots n_{d},t}$ as the probability that, at a given time, $n_1$ individuals of type 1, $n_2$ individuals of type 2 are present and so on. At time $t$ the system is updated as follows:
\begin{equation}
P_{\bm{n},t+1}=\sum_{\bm{m}}Q_{\bm{n};\bm{m}}P_{\bm{m},t}.
\label{chain2D}
\end{equation}
The $d$-dimensional matrix $\bm{Q}$ is a natural generalization of the one-dimensional case (see Eq.~\eqref{binomial}) with $\bm{Q}$ having a $(d+1)$-nomial structure~\cite{hogg_13}
\begin{equation}
Q_{\bm{n};\bm{m}}=\frac{N!}{n_1!n_2!\hdots n_d!(N-\sum_{i=1}^d n_i)!}p_1^{n_1}p_2^{n_2}\hdots p_d^{n_d}\left(1-\sum_{i=1}^{d} p_i\right)^{N-\sum_{i=1}^d n_i},
\label{trinomial}
\end{equation}
where, in a single multinomial trial, $p_i$ is the probability of event $i$ being observed. The probability distribution function (pdf) is a $(d+1)$-nomial, rather than a $d$-nomial, since there is (as in the 1D case~\cite{challenger_13,challenger_14}) formally a $(d+1)^{\rm th}$ species of population $N - \sum^{d}_{i=1} n_i$, which represents the free capacity of the system.

The probabilities $p_i$ will have the general form
\begin{equation}
(p_1,p_2,\hdots,p_d)=\bm{p}\left( \frac{\bm{m}}{N} \right),
\label{f1f2}
\end{equation}
where $\bm{m}$ is the initial state of the system. The functions $\bm{p}$ will define the model; some of the forms considered in past work are given in Sec.~\ref{sec:lv}. We note here that, as we are dealing with probabilities, we require that
\begin{equation}
\sum^{d}_{i=1} p_i\left( \frac{\bm{m}}{N}\right) < 1,
\label{less_than_one}
\end{equation}
for all allowed combinations of $\bm{m}$. One consequence of moving to multiple dimensions is that the transition matrix becomes quite cumbersome to implement numerically, even for moderate values of $N$. For example, in 2D the transition matrix will contain $(N+1)^4$ entries. It therefore becomes even more important to have a mesoscopic description of the system available, where $N$ is simply a parameter, and so the analysis does not become unfeasible for moderate or large values of $N$. Below we outline the ideas behind the derivation of this mesoscopic description. The method is similar to the 1D case~\cite{challenger_14}, and the details of the derivation are given in Appendix~\ref{sec:2-d-dim}. 

Our starting point is the assumption that the system is described by the state variables $z_i = n_i/N$, $i=1,\hdots,d$. For large $N$, a good approximation is to assume that the $z_i$ are continuous; this is the key assumption in the mesoscopic modeling of the system. If in addition we assume that the system is Markovian, then it will satisfy the Chapman-Kolmogorov equation~\cite{kampen_07}:
\begin{equation}
\label{CK}
P(\bm{z},t+1 \mid \bm{z}_0,t_0) = \int \textrm{d} \bm{z}' P(\bm{z},t+1 \mid \bm{z}',t) P(\bm{z}',t \mid \bm{z}_0,t_0),
\end{equation}
where $\bm{z}=(z_1,\hdots,z_d)$ (note that we have temporarily represented time as an argument, rather than a subscript, for clarity). To obtain a Fokker-Planck-like equation, one introduces a change of variable $z_i' = z_i- \Delta z_i$, and rewrites the above equation using a Taylor expansion. The properties of the system are now represented by jump moments $M_{\bm{\ell}}(\bm{z})$ defined by
\begin{equation}
M_{\bm{\ell}}(\bm{z}) = \left\langle \left(z_{1,t+1}-z_{1,t}\right)^{\ell_1} \hdots \left(z_{d,t+1}-z_{d,t}\right)^{\ell_d}  \right\rangle_{\mathbf{z}_t = \mathbf{z}},
\end{equation}
where $\bm{\ell}$ is a $d$-dimensional vector of integers (see Appendix~\ref{sec:2-d-dim}). 

However, unlike the standard case, where time is continuous~\cite{gardiner_09}, we are unable to neglect the higher-order jump moments. This is because, from one time step to the next, large changes in $\bm{z}$ are possible; in the continuous time case the jump moments were of order $dt$, and were by definition small. Instead, as in the 1D case~\cite{challenger_14}, we use the fact that we expect jumps from $p_{i,t}$ to $z_{i,t+1}$ to be small, and so work with a new set of jump moments defined by 
\begin{equation}
\label{J}
 J_{\bm{r}}(\bm{p}) = \left\langle \left(z_{1,t+1}-p_{1,t}\right)^{r_1} \hdots \left(z_{d,t+1}-p_{d,t}\right)^{r_d} \right\rangle_{\mathbf{z}_t = \mathbf{z}},
\end{equation}
where $\bm{r}$ is another $d$-dimensional vector of integers and where $p_{i,t} \equiv p_i(\bm{z}_t)$. The higher-order $J_{\bm{r}}(\bm{p})$, unlike the $M_{\bm{\ell}}(\bm{z})$, can be shown to be of order $N^{-2}$ (see Appendix~\ref{sec:j_d}), and so may be neglected. This means that the Taylor expansion may be truncated, giving a second-order differential equation in the variables $z_i$. A consequence of working with the jump moments $J_{\bm{r}}$ rather than $M_{\bm{\ell}}$ is the appearance of pdfs $\mathcal{P}_t(\bm{p})$, defined such that
\begin{equation}
P_t(\bm{z})=\mathcal{P}_t(\bm{p})|\textrm{det}(\bm{J})|\,\,\,\, \textrm{with}\,\,\, J_{ij}=\frac{\partial p_i}{\partial z_j}.
\label{jac_trans}
\end{equation}
In Appendix~\ref{sec:2-d-dim} it is shown that for large $N$ Eq.~\eqref{CK} may be written as
\begin{equation}
\label{J_final}
P_{t+1}(\mathbf{z}) =  \mathcal{P}_t(\mathbf{z}) + \frac{1}{4 N } \sum_{i,j=1}^d \frac{\partial^2}{\partial z_i \partial z_j} \left[ \left(z_i(\delta_{ij}-z_j)+ z_j(\delta_{ij}-z_i)\right) \mathcal{P}_t(\mathbf{z}) \right],
\end{equation}
where we have reverted to using a subscript for $t$ and have dropped the dependence on initial conditions (so that $P(\bm{z},t \mid \bm{z}_0,t_0)$ now reads $P_t(\bm{z})$). This is the equivalent equation to the Fokker-Planck equation found in the mesoscopic description of continuous-time Markov processes.

Just as Fokker-Planck equations can be shown to be equivalent to stochastic differential equations~\cite{gardiner_09}, in Appendix~\ref{sec:2-d-dim}, we show that Eq.~\eqref{J_final} is equivalent to the stochastic difference equation
\begin{equation}
\bm{z}_{t+1}=\bm{p}(\bm{z_t}) + \bm{\eta}_t = \bm{p}_t +\bm{\eta}_{t}.
\label{sde}
\end{equation}
Here $\bm{\eta} = (\eta_1,\hdots,\eta_d)$ is a Gaussian noise with zero mean and correlator 
\begin{equation}
\langle \eta_{i,t} \eta_{j,t'} \rangle = \frac{1}{N}\,B_{ij}\,\delta_{tt'},
\label{sde_correlator}
\end{equation}
where $B_{ii}=p_i(1-p_i)$ and $B_{ij}=-p_i p_j$, if $i \neq j$. 

We emphasize again that Eq.~\eqref{sde} is not simply the deterministic map with added noise; the noise is derived from the microscopic model and has a form dictated by that model. It is multiplicative --- that is, it depends on the dynamical variable --- and is not simply additive. Equations (\ref{J_final}) and (\ref{sde}) give equivalent mesoscopic descriptions of the system, and will form the starting points for the analysis that we will present in this paper.

\subsection*{The linear theory}

In the previous study of the 1D case~\cite{challenger_14}, we showed that approximate analytic results for the fluctuations around a non-chaotic attractor could be obtained by linearizing around the attractor of the deterministic map, \emph{e.g.} a fixed point, which we denote by $z^*$. We can extend this to higher dimensions by making an analogous substitution $\bm{z}_t=\bm{z}^*+\bm{\xi}_t/\sqrt{N}$ in Eq.~\eqref{sde}. Equating terms of the same order in $N$ we obtain
\begin{equation}
\bm{\xi}_{t+1}=\bm{J}^*\bm{\xi}_{t}+\bm{\rho}_{t},
\label{linear_d_e}
\end{equation}
where the Jacobian has been evaluated at the fixed point and the noise term $\bm{\rho}_t$ has the same correlator as $\bm{\eta}$ (but without the factor $1/N$), again evaluated at the fixed point. We will denote this correlator by $\bm{B}^*$. Successive applications of the linear map, initialized at $\bm{\xi}_0$ at time $t_0$ lead us to
\begin{equation}
\bm{\xi}_t=(\bm{J}^*)^{t-t_0}\bm{\xi}_0 +\sum_{m=0}^{t-(t_0+1)}(\bm{J}^*)^{m}\bm{\rho}_{t-(m+1)}.
\label{soln_to_lin}
\end{equation}
To lighten the notation, we will drop the asterisk from $\bm{J}^*$ and $\bm{B}^*$, assuming that all quantities are evaluated at the fixed point. The quantities of interest for us are the first two moments, which can be written as
\begin{eqnarray}
\langle \bm{\xi}_t\rangle&=&(\bm{J})^{t-t_0}\bm{\xi}_0\,,\nonumber\\
\langle \bm{\xi}_t\bm{\xi}^T_t\rangle&=&(\bm{J})^{t-t_0}\bm{\xi}_0\bm{\xi}_0^T(\bm{J}^T)^{t-t_0}+
\left\langle\sum_{m=0}^{t-(t_0+1)} (\bm{J})^m\bm{\rho}_{t-(m+1)}
\sum_{m'=0}^{t-(t_0+1)}\bm{\rho}_{t-(m'+1)}^T(\bm{J}^{m'})^T\right\rangle \nonumber \\
&=&(\bm{J})^{t-t_0}\bm{\xi}_0\bm{\xi}_0^T(\bm{J}^T)^{t-t_0}+
\sum_{m=0}^{t-(t_0+1)}\bm{J}^m\bm{B}(\bm{J}^{m})^T.
\end{eqnarray}

For both the first and second moment, the first term in the above expressions provides information about the initial condition. This information will be lost taking $t_0\to -\infty$. This means that, in the stationary state, the first moment is zero. To calculate the variance of the fluctuations in the stationary state, it is useful to consider the diagonalization of $\bm{J}$ \emph{i.e.} $\bm{J}=\bm{P}\bm{D}\bm{P}^{-1}$, where $\bm{D}$ is diagonal---with the eigenvalues of $\bm{J}$, say $\lambda$, as its entries. The matrix $\bm{P}$ is constructed from the eigenvectors of $\bm{J}$. We will also use the fact that $(\bm{P}\bm{D}\bm{P}^{-1})^l=\bm{P}\bm{D}^l\bm{P}^{-1}$. Therefore, the stationary covariance matrix, $\bm{\Xi}$, can be written as (stationary moments being denoted by the superscript ``st'')
\begin{equation}
\bm{\Xi}=\langle \bm{\xi}\bm{\xi}^T\rangle^{\textrm{st}}=
\sum_{m=0}^{\infty}\bm{P}\bm{D}^{m}\bm{P}^{-1}\bm{B}(\bm{P}^{-1})^T\bm{D}^m\bm{P}^T,
\end{equation}
or, introducing the matrix $\bm{C}=\bm{P}^{-1}\bm{B}(\bm{P}^{-1})^T$,
\begin{equation}
\bm{P}^{-1}\bm{\Xi}(\bm{P}^T)^{-1}=\sum_{m=0}^{\infty}\bm{D}^m\bm{C}\bm{D}^m.
\label{sum_on_C}
\end{equation}

Now the $ij^{\rm th}$ entry of the matrix $\bm{D}^m\bm{C}\bm{D}^m$ is $c_{ij}\left( \lambda_{i} \lambda_{j} \right)^m$, where $c_{ij}$ are the entries of the matrix $C$. Therefore the sum on $m$ in Eq.~\eqref{sum_on_C} is a geometric sum and can be performed as long as $| \lambda_i | < 1$ (recall that the eigenvalues of $\bm{J}$ will in general be complex). However, if the fixed point is stable, this will be the case. Carrying out the sum, we obtain
\begin{equation}
\left[ \bm{P}^{-1}\bm{\Xi}(\bm{P}^T)^{-1} \right]_{ij} = \frac{c_{ij}}{1 - \lambda_{i}\lambda_{j}},
\label{C_sum}
\end{equation}
and so
\begin{equation}
\Xi_{ij} = \sum_{k,\ell} P_{ik}\,\frac{c_{k\ell}}{1 - \lambda_{k}\lambda_{\ell}}\,
P_{j\ell} = \sum_{k,\ell,r,s} \frac{P_{ik} P^{-1}_{kr} B_{r s} P^{-1}_{\ell s} P_{j \ell}}{1 - \lambda_{k}\lambda_{\ell}}.
\label{linearise_fp}
\end{equation}

The analysis can be extended to the case of an $n$-cycle. Again one performs a linearization, but this time around the $n$ points that comprise the cycle, say $\bm{z}_1,\bm{z}_2,\hdots \bm{z}_n$, $a=1,\hdots,n$. If we start close to the point $\bm{z}_a$, iterate the map $n$ times, and equate terms of the same order in $N$, one finds in an exactly analogous way to the 1D case \cite{challenger_14},
\begin{equation}
\bm{\xi}_{t+n}^{(a)}=\bm{J}^{(a)}\bm{\xi}_{t}^{(a)}+\bm{\sigma}_t^{(a)}.
\label{sigma_eqn}
\end{equation}
Here $\bm{J}^{(a)}=\bm{J}_{a-1}\bm{J}_{a-2}\hdots \bm{J}_{a-n}$ and 
\begin{equation}
\bm{\sigma}^{(a)}_t=\bm{\rho}_{t+n-1}^{(a-1)}+\sum\limits_{m=2}^n \bm{A}_{a-(m-1)}\bm{\rho}_{t+n-m}^{(a-m)},
\end{equation}
where the labels are to be taken as $\textrm{mod}n$ and $\bm{A}_{a-i}=\bm{J}_{a-1}\bm{J}_{a-2}\hdots\bm{J}_{a-i}$. 

Equation \eqref{sigma_eqn} has exactly the same form as the original difference equation \eqref{linear_d_e}, except that $t$ changes by $n$ at every iteration. So we can proceed as in the case of a fixed point, and find the first two moments in the stationary state, where knowledge of the initial condition has been lost. Using the results found above, along with the periodicity of the system, we can write (compare with Eq.~\eqref{soln_to_lin})
\begin{equation}
\bm{\xi}_t^{(a)}=\sum_{m=0}^{\infty}(\bm{J}^{(a)})^m\bm{\sigma}^{(a)}_{t-n(m+1)}.
\end{equation}
This implies that in the stationary state the first order moments are equal to zero, and that the second order stationary moments are given by
\begin{equation}
\langle\bm{\xi}\bm{\xi}^T\rangle_a^{\rm st}=\sum_{m=0}^{\infty}\bm{J}^{(a)^m}\bm{B}^{(a)}(\bm{J}^{(a)^T})^m,
\end{equation}
where 
\begin{equation}
\bm{B}^{(a)} = \bm{B}_{a-1} + \sum\limits_{m=1}^{n-1}\bm{A}_{a-m}\mathbf{B}_{a-(m+1)} \bm{A}_{a-m}^{T}.
\end{equation}
We now write $\bm{J}^{(a)}=\bm{V}^{(a)}\bm{\Lambda}^{(a)}(\bm{V}^{(a)})^{-1}$, where $\bm{V}^{(a)}$ is the matrix which diagonalizes $\bm{J}^{(a)}$, and $\bm{\Lambda}^{(a)}$ is the diagonal matrix of its eigenvalues. We are able to calculate the stationary covariances, $\Xi_{ij}$, just as we did for the fixed point. One finds that
\begin{equation}
\Xi_{ij}^{(a)} = \sum_{k,\ell,r,s} \frac{V^{(a)}_{ik} \left( V^{(a)}\right)^{-1}_{kr} B^{(a)}_{r s} \left( V^{(a)} \right)^{-1}_{\ell s} V^{(a)}_{j \ell}}{1 - \lambda^{(a)}_{k}\lambda^{(a)}_{\ell}}.
\label{linearise_n_cycle}
\end{equation}

Finally, we can perform a similar analysis on the quasiperiodic behavior, which is not seen in the 1D case. The difference between quasiperiodic and periodic ($n$-cycle) behavior can be understood by visualizing the evolution of a 2D map in the $(x,y)$ plane. In the latter case, once the transient behavior has been discarded, it is found that only $n$ points are visited, always in the same order, and so the motion is exactly periodic. In the quasiperiodic case, although the motion may appear to be periodic from the time series, in the $(x,y)$ plane the map never quite returns to a point previously visited. If the map is iterated for long enough, the points are sufficiently close to each other that they appear to form a closed ring, as we shall see in Section~\ref{sec:lv}. To proceed with the linearization, we write $\bm{z}_t=\bm{z}_t^{*}+\bm{\xi}_t/\sqrt{N}$, where $\bm{z}^*$ is now the deterministic, quasiperiodic trajectory. 
\begin{equation}
\bm{\xi}_{t+1}=\bm{J}(\bm{z}_t^*)\bm{\xi}_t+\bm{\rho}_{t},
\end{equation}
with
\begin{equation}
\langle \rho_{i,t} \rho_{j,t'} \rangle=B_{ij}(\bm{p}(\bm{z}^*_t))\,\delta_{tt'}.
\end{equation}
We will use the initial condition $\bm{\xi}_{t=0}=\bm{\xi}_0$. Then
\begin{equation}
\bm{\xi}_t=\bm{\tilde{J}}^{t}_0\bm{\xi}_0+\sum_{m=1}^{t-1}\bm{\tilde{J}}^{t}_{t-m}\bm{\rho}_{t-(m+1)}+\bm{\rho}_{t-1},
\label{soln_QP}
\end{equation}
where we have introduced $\bm{\tilde{J}}^b_a=\bm{J}(\bm{z}_{b-1}^*)\bm{J}(\bm{z}_{b-2}^*)\hdots \bm{J}(\bm{z}_a^*)$. From Eq.~\eqref{soln_QP}
\begin{eqnarray}
\langle \bm{\xi}_t \rangle=&\bm{\tilde{J}}^{t}_0&\bm{\xi}_0 \nonumber\\
\langle \bm{\xi}_t \bm{\xi}_t^T\rangle=&\bm{\tilde{J}}^{t}_0&\bm{\xi}_0\bm{\xi}_0^T(\bm{\tilde{J}}_0^{t})^T
+\sum_{m=1}^{t-1}\bm{\tilde{J}}^{t}_{t-m}\bm{B}(\bm{p}(\bm{z}^*_{t-(m+1)}))
(\bm{\tilde{J}}^{t}_{t-m})^T+\bm{B}(\bm{p}(\bm{z}^*_{t-1})).
\label{linearise_qp}
\end{eqnarray}
Notice that, unlike the cases of the fixed point or $n$-cycle, these are not the stationary moments. This is because there is a neutrally stable direction for fluctuations in the quasiperiodic case \cite{kaneko_86}, which means that the fluctuations can grow with time. Consequently, the moments must be calculated using an initial condition and at a particular time, $t$. At very large times, our predictions will begin to lose accuracy, as the fluctuations reach the size of the attractor.

\subsection*{The power spectra}

Unlike the 1D case, in 2D maps the eigenvalues of the Jacobian, evaluated at the fixed point, can be complex. If this is the case, then the fixed point will be approached \emph{via} damped oscillations. Using the formalism developed above, we will show that the stochasticity present in the finite-$N$ system is able to sustain the oscillations, which means that they do not die out. Similar results are well-known for continuous time systems, where the sustained oscillations are characterized by the power spectra~\cite{mckane_05}. Here we report similar results for discrete-time models, although the power spectrum has a different analytical form. In this sub-section we will restrict ourselves to 2D, although much of the discussion holds for a general number of dimensions, $d$.

Our starting point for the theoretical description of the oscillations will be the linearized difference equation, given in Eq~\eqref{linear_d_e}. We will take the Fourier transform of this equation, and find a closed form expression for $\langle |\tilde{\xi}_i(\omega)|^2\rangle$, for $i=1,2$. To do this, we will need to use the discrete-time Fourier transform \cite{marple_87}: 
\begin{equation}
\bm{\tilde{\xi}}(\omega) = \sum^{\infty}_{t = - \infty} \bm{\xi}_{t}\,e^{-i\omega t},
\label{FT_def}
\end{equation}
where the sum is over integers $t$. Since $\bm{\tilde{\xi}}(\omega)$ is a periodic function of $\omega$ with period $2\pi$, we restrict $\omega$ so that $0 \leq \omega < 2\pi$. Taking the Fourier transform of the linearized difference equation we find
\begin{equation}
\mathrm{e}^{i\omega}\bm{\tilde{\xi}}(\omega)=\bm{J}\bm{\tilde{\xi}}(\omega)+\tilde{\bm{\rho}}(\omega).
\end{equation}
Omitting the $\omega$-dependence of $\bm{\tilde{\xi}}$ and $\bm{\tilde{\rho}}$,
and introducing the $2 \times 2$ unit matrix $\bm{I}$, we may write
\begin{equation}
(\mathrm{e}^{i\omega}\bm{I}-\bm{J})\tilde{\bm{\xi}}=\tilde{\bm{\rho}}, \,\,\,\, \textrm{and so}\,\,\,\, \tilde{\bm{\xi}} =(\mathrm{e}^{i\omega}\bm{I}-\bm{J})^{-1}\tilde{\bm{\rho}}.
\end{equation}

Using the same notation as in the continuous time case~\cite{dauxois_09}, we define the matrix $\bm{\Phi}$ to be $\mathrm{e}^{i\omega}\bm{I}-\bm{J}$, so that $\tilde{\bm{\xi}} =\bm{\Phi}^{-1}\tilde{\bm{\rho}}$. Therefore
\begin{equation}
\tilde{\bm{\xi}}\tilde{\bm{\xi}}^{\dagger}=\bm{\Phi}^{-1} \tilde{\bm{\rho}}\,\tilde{\bm{\rho}}^{\dagger}\left( \bm{\Phi}^{\dagger} \right)^{-1}.
\end{equation}
Taking the expectation of the above expression and rewriting in component form leads to 
\begin{equation}
P_i(\omega)=\langle |\tilde{\xi}_i(\omega)|^2 \rangle=\sum_{j=1}^2\sum_{k=1}^2 \Phi^{-1}_{ij}(\omega)B_{jk}\left(\Phi^{\dagger}_{ki}(\omega)\right)^{-1},
\label{power_spectra}
\end{equation}
where we are only considering the diagonal entries of the power spectral density matrix and where $\bm{B}$ is the noise-correlator matrix. 

As for the continuous-time case, we wish to know whether $P_i(\omega)$ has peaks for non-zero $\omega$ and if so, the value of these $\omega$. A first approximation may be found by finding the value of $\omega$ for which the denominator of the power spectrum, $D(\omega) = \mathrm{det}\,\Phi(\omega)\Phi^{\dagger}(\omega)$ is a minimum. For the case $d=2$,
\begin{equation}
D(\omega)=\prod_{j=1}^2|\lambda_j-\mathrm{e}^{i\omega}|^2,
\end{equation}
and since we are interested in cases where the eigenvalues are complex, they will form a conjugate pair. Writing the eigenvalues in terms of their magnitude and complex phase \emph{i.e.} $(\lambda_1,\lambda_2)=(|\lambda|\mathrm{e}^{i\theta}, |\lambda|\mathrm{e}^{-i\theta})$, $D(\omega)$ becomes
\begin{equation}
D(\omega)= \left[ |\lambda|^2-2|\lambda|\mathrm{cos}(\omega-\theta)+1\right]
\left[ |\lambda|^2-2|\lambda|\mathrm{cos}(\omega+\theta)+1\right].
\label{D_of_omega}
\end{equation}
Both terms in the square brackets are positive for $|\lambda| < 1$, and we can obtain a crude estimate of the position of the peaks by asking that one or other of these terms are minimized. This occurs when $\mathrm{cos}(\omega \pm \theta)=1$, or when $\omega=\theta, 2\pi - \theta$. A more refined estimate is obtained by minimizing the entire expression for $D(\omega)$ to obtain
\begin{equation}
\cos\omega = \frac{\left( |\lambda|^2 + 1 \right)}{2|\lambda|}\,\cos\theta.
\label{peaks}
\end{equation}
So, it is the phase of the complex eigenvalues which largely influences the position of the peaks in the power spectrum. This differs from the continuous-time case, where it is the complex part of the eigenvalue which determines the position of the peak~\cite{mckane_05}. However, if one considers the discrete-time deterministic model near the fixed point (Eq.~\eqref{soln_to_lin}, but with $\bm{\rho}$ set equal to zero), then it is the phase which determines the frequency of the damped oscillations observed as a trajectory approaches the fixed point.

\section{The Lotka-Volterra model}
\label{sec:lv}

To illustrate the theory developed in the previous section we present a 2D system, which we will describe using the Markov chain model. As pointed out in the Introduction, there is a wide range of models in the literature representing two-species interactions. Those corresponding to competition between species may be written in the general form~\cite{hassell_76,may_76,sole_92c}:
\begin{align}
x_{t+1}=x_t [g(x_t+\alpha y_t)]^{-b_1},\,\,\, y_{t+1}=y_t [h(y_t+\beta x_t)]^{-b_2},
\end{align}
where $g$ and $h$ are functions and $\alpha, \beta, b_1$ and $b_2$ are constants. Models of this kind have been studied extensively~\cite{hassell_76,may_76,sole_92c}, and in all cases it is found that they follow the period-doubling route to chaos. This is not of such great interest to us in this paper, since this type of behavior is already present in one-dimensional systems. On the other hand, predator-prey and host-parasitoid models display a wider variety of behaviors. There exist several classes of models, most of which can be written in one of the following forms~\cite{beddington_75,beddington_76,lauwerier_86a,lauwerier_86b,may_87,sole_92b}:
\begin{eqnarray}
&x_{t+1}=x_t \exp [r(1-x_t)-a y_t)],\,\,\, &y_{t+1}=c x_t [1-\exp(-a y_t)],\\
&x_{t+1}=a x_t\psi (y_t),\hspace*{6.7em} &y_{t+1}=ax_t-x_{t+1},\\
&x_{t+1}=\mu x_t \phi(x_t+y_t),\hspace*{4.5em} &y_{t+1}=\beta x_t y_t,\label{lvgen}\\
&x_{t+1}=\lambda x_t[1+y_t]^{-k},\hspace*{4.5em} &y_{t+1}=x_ty_t[1+y_t]^{-k-1},
\end{eqnarray}
where $\phi$ and $\psi$ are functions and where all other parameters are constants. These models, where $x$ and $y$ represent the size of the prey/host and predator/parasitoid populations, respectively, generally undergo a Hopf bifurcation from a stable fixed point, giving rise to periodic or quasiperiodic states. These then bifurcate into a series of higher period orbits---intermingled with quasiperiodic bands in the latter case---until eventually entering a chaotic region. We shall focus, then, on these types of interactions.

\begin{figure}
\begin{center}
\includegraphics[scale=1.9]{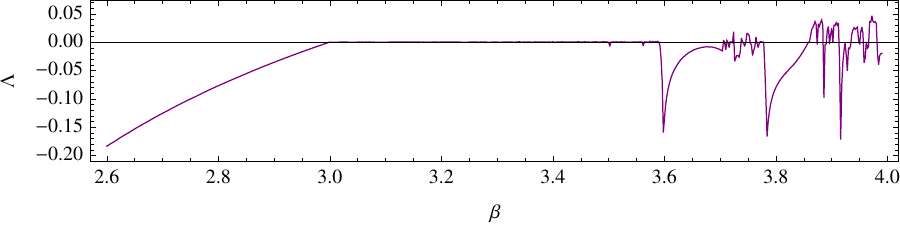}
\end{center}
\caption{Largest Lyapunov exponent of the Lotka-Volterra map, as the parameter $\beta$ is varied, fixing $\mu=3$.}
\label{fig:lyap_vs_beta}
\end{figure}

The simplest model displaying all of the features we are interested in is that in Eq.~\eqref{lvgen} which, taking $\phi(\xi)=1-\xi$, has the Lotka-Volterra form~\cite{lauwerier_86a,sole_92b}:
\begin{eqnarray}
x_{t+1}&=&\mu x_t(1-x_t-y_t),\label{lvx} \\
y_{t+1}&=&\beta x_ty_t.\label{lvy}
\end{eqnarray}
In this model, the prey population grows logistically, and it is depleted by the presence of predators in the system; these, in turn, increase in number with the abundance of prey---$\mu$ and $\beta$ are real parameters, which we take to lie in the range $(0,4)$. Varying the parameter $\beta$ for a given value of $\mu$, we can explore a range of different behaviors. Figure~\ref{fig:lyap_vs_beta} shows the largest Lyapunov exponent ($\Lambda$) of the deterministic map \cite{ott_93} over a range of $\beta$, fixing $\mu=3$. Moving from left to right, we start with a stable fixed point. The fixed point loses stability at $\beta=3$, as the system undergoes a Hopf bifurcation and becomes quasiperiodic, indicated by a Lyapunov exponent equal to zero \cite{kaneko_86,hilborn_2000}. Periodic windows appear, which are shown by a negative value of $\Lambda$. Transitions to chaos may occur from these periodic windows \cite{kaneko_86}.

We will look at the effects of intrinsic noise on these different attractors in turn. We shall, therefore, use Eqs.~\eqref{lvx}--\eqref{lvy} to motivate our choice for the microscopic probabilities $p_i$ in the Markov chain model. In Eqs.~\eqref{trinomial} and \eqref{f1f2}, with $d=2$, we introduce
\begin{equation}
p_1=\mu\frac{m_1}{N}\frac{N-m_1-m_2}{N},\,\,\, p_2=\beta\frac{m_1}{N}\frac{m_2}{N}.
\end{equation}
Note that $p_1$ and $p_2$, as defined above, satisfy the inequality in \eqref{less_than_one}, provided $0\leq m_{1},m_{2} \leq N$. In the thermodynamic limit, $N \to \infty$, we recover the Lotka-Volterra map Eqs.~\eqref{lvx} and \eqref{lvy}.

\begin{figure}
\includegraphics[scale=1]{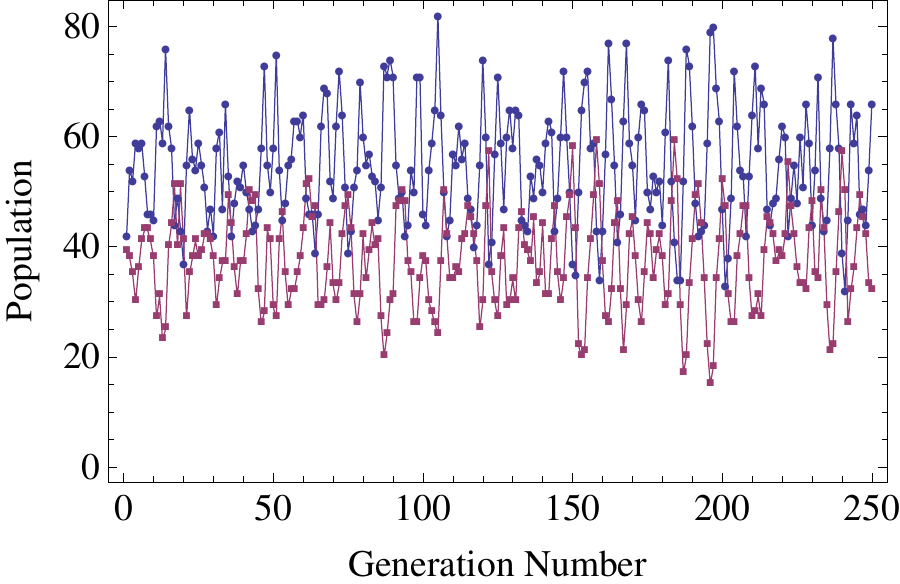}
\caption{A single stochastic realization from the Markov chain model of the Lotka-Volterra system, showing species $x$ (blue) and $y$ (purple). Parameters used were $N=140$, $\mu=3$ and $\beta=2.6$.}
\label{fig:WF_sim}
\end{figure}

Figure~\ref{fig:WF_sim} shows one realization of this process, found by simulating the Markov chain for $N=140$. However, we shall mainly use the stochastic difference equations to obtain results in this section, due to the difficulty in using the Markov chain for larger values of $N$. The stochastic difference equations are defined in Eqs.~\eqref{sde} and \eqref{sde_correlator} with this choice of the functions $p_1$ and $p_2$. 


\begin{figure}
\begin{center}
\subfigure[]{\includegraphics[scale=0.96]{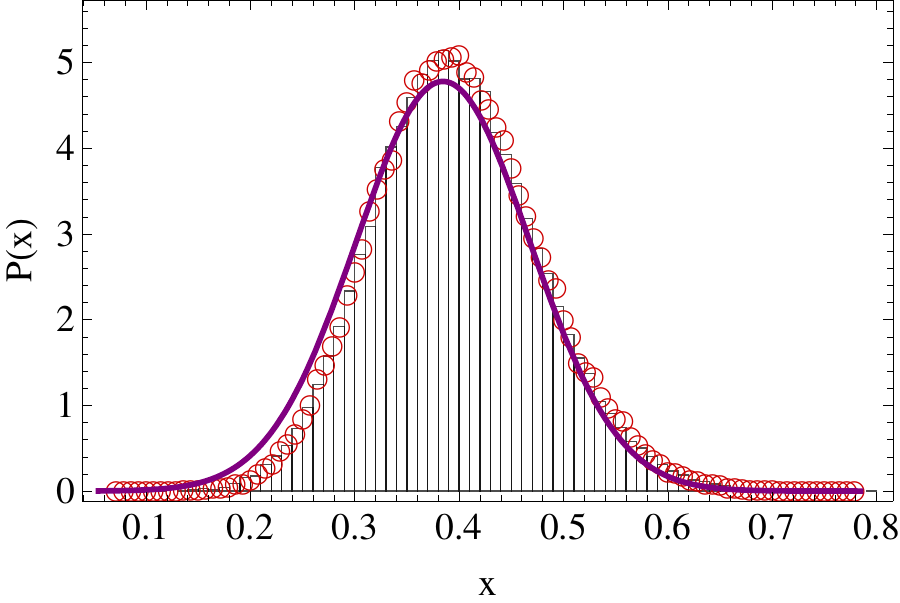}}
\subfigure[]{\includegraphics[scale=0.96]{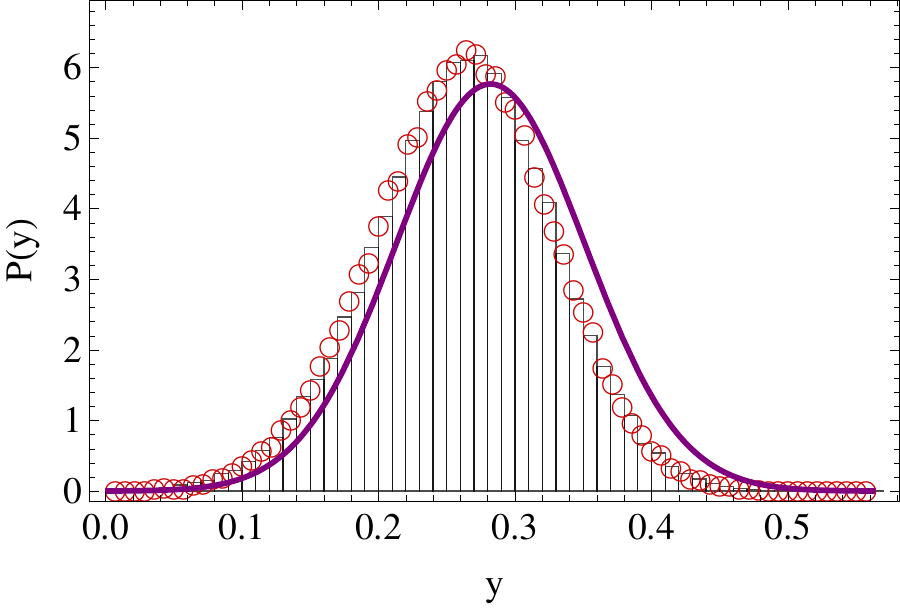}}
\end{center}
\caption{Comparison of simulation data obtained with the Markov chain (red circles), and data obtained from the stochastic difference equation (bars) for the Lotka-Volterra model. Parameters used were $N=140$, $\mu=3$ and $\beta=2.6$. The theoretical predictions, found from linearizing the difference equation around the fixed point, is shown by the purple curve.}
\label{fig:comp_N140}
\end{figure}

\begin{figure}
\begin{center}
\subfigure[]{\includegraphics[scale=0.96]{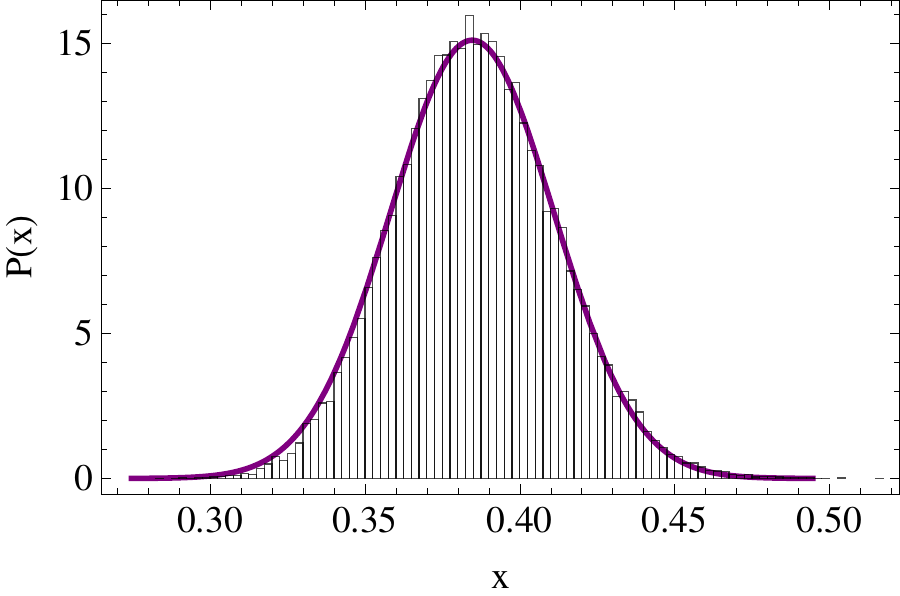}}
\subfigure[]{\includegraphics[scale=0.96]{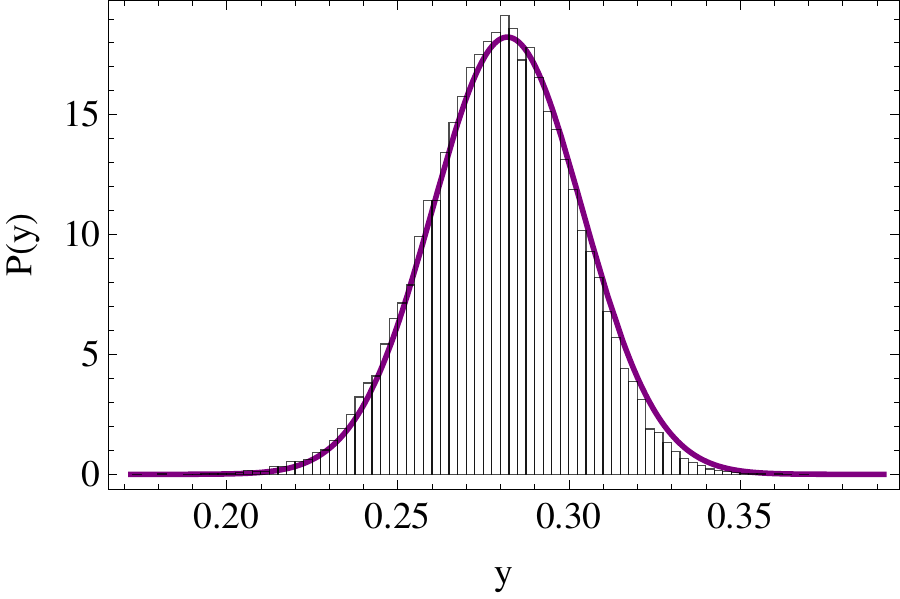}}
\end{center}
\caption{Comparison between simulation data from the stochastic difference equation \eqref{sde} (bars) and the analytical prediction given by Eq.~\eqref{linearise_fp} (purple curve), obtained by linearizing the equations around the fixed point. The parameters used for the Lotka-Volterra model were $\mu=3$ and $\beta=2.6$, and $N=1400$.}
\label{fig:comp_N1400}
\end{figure}

We will begin by looking at the fluctuations around the stable fixed point, using the ideas developed in the previous section. Figure~\ref{fig:comp_N140} compares the simulation results from the Markov chain with those from the stochastic difference equations for $N=140$, showing good agreement. The histograms showing results from the stochastic difference equations were obtained using 80,000 data points. An error analysis was carried out, using an ensemble of 100 stochastic simulations. The variation of the height of each bin was very small: less than the diameter of the circles used to show the results from the Markov chain. Therefore we do not display the error bars in the histograms, here or elsewhere in the article. This figure also displays the probability distribution predicted by the linear theory found in the previous section. At this value of $N$ the linear theory is not very accurate: in particular, it cannot capture the skewness of the distribution that is apparent at small $N$. However, the theory is much more accurate at larger $N$. Figure~\ref{fig:comp_N1400} shows very good agreement between the nonlinear stochastic difference equation and the theoretical prediction for $N=1400$.


\subsection*{Quasi-cycles}

\begin{figure}
\begin{center}
\subfigure[]{\includegraphics[scale=0.96]{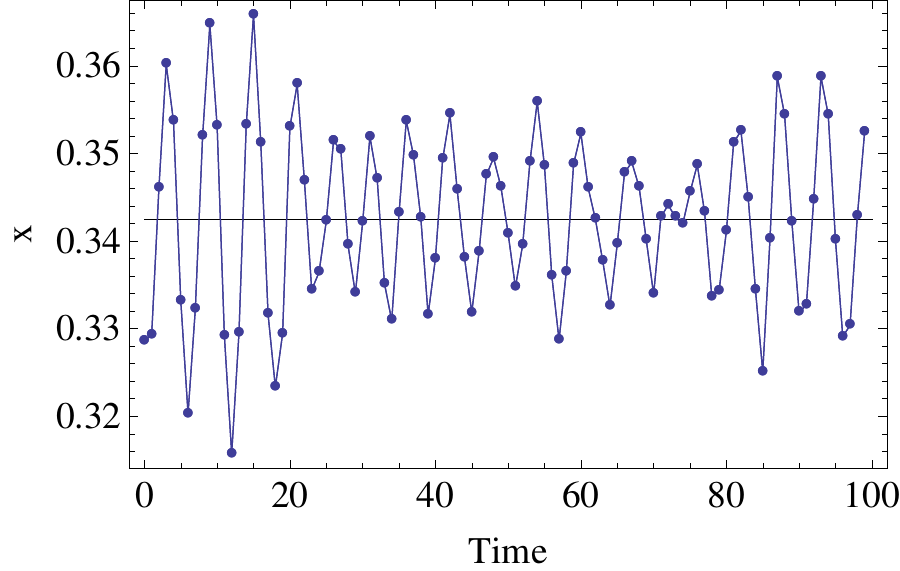}}
\subfigure[]{\includegraphics[scale=0.96]{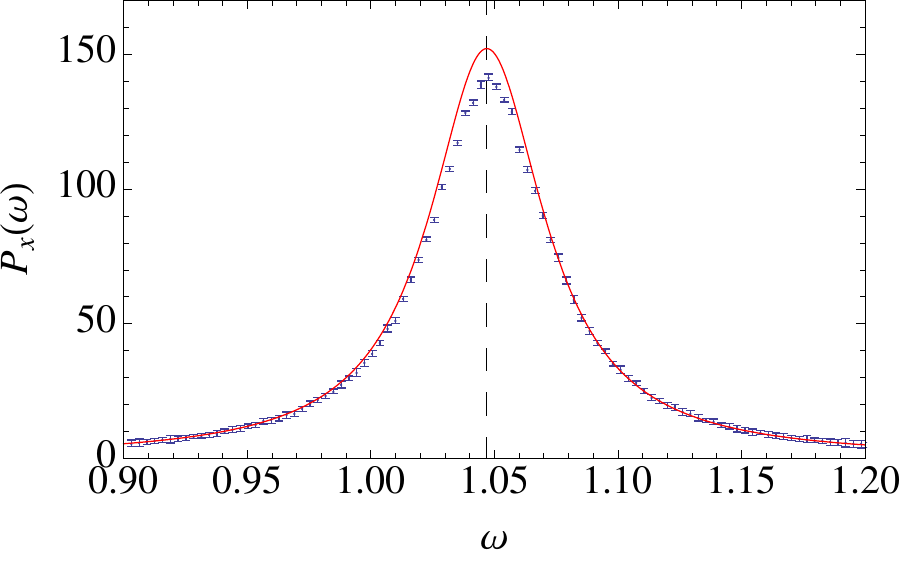}}
\end{center}
\caption{(a) Stochastic oscillations in the concentration of species $x$ in the Lotka-Volterra map. Parameters values are $N=50000$, $\mu=3$ and $\beta=2.92$. The black line indicates the value of the fixed point for $x$ in the deterministic map. (b) The power spectrum of species $x$ from the theory (red line) and simulation (blue dots with error bars), averaged over 4000 realizations. The vertical dashed line locates the complex phase of the eigenvalues.}
\label{fig:quasicycles_x}
\end{figure}

\begin{figure}
\begin{center}
\subfigure[]{\includegraphics[scale=0.96]{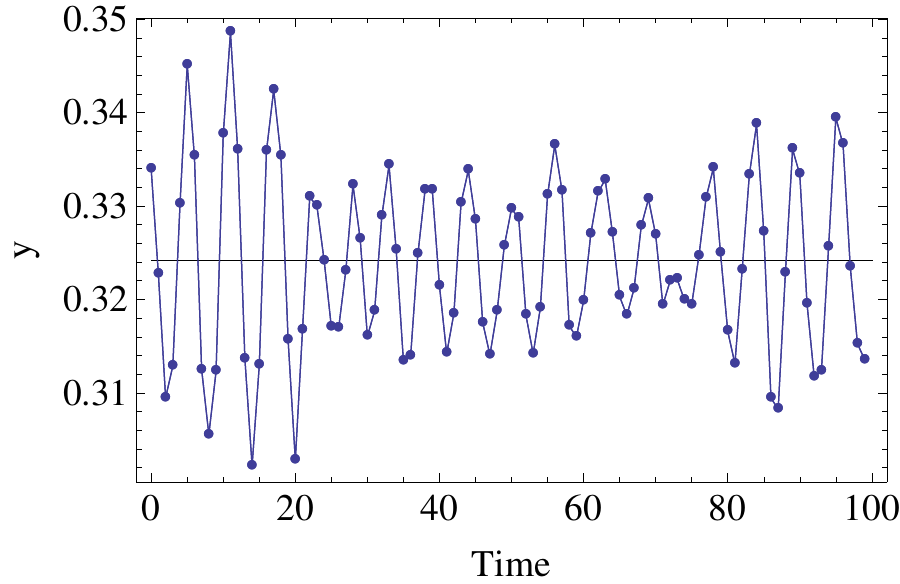}}
\subfigure[]{\includegraphics[scale=0.96]{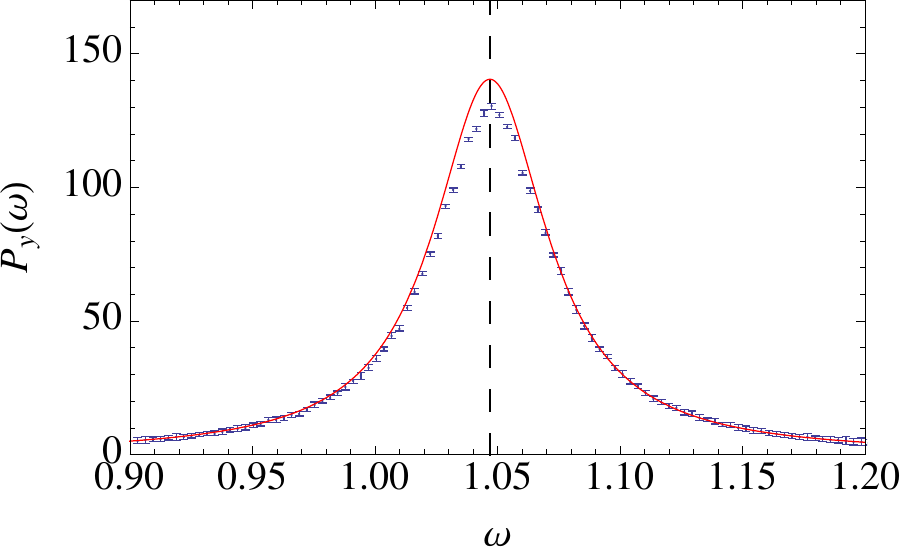}}
\end{center}
\caption{Same as in Fig. \ref{fig:quasicycles_x}, but for $y$.}
\label{fig:quasicycles_y}
\end{figure}

We now show some results for a choice of parameters where quasi-cycles are visible. By inspecting the form of the power spectrum, derived in the previous section, it is clear that the frequency dependence enters through terms of the form $\textrm{e}^{i\omega}$. Therefore, the power spectrum is automatically periodic in $2\pi$. Due to the even nature of the function, all information is contained in the range $0\leq \omega <\pi$. In Figures~\ref{fig:quasicycles_x} and \ref{fig:quasicycles_y} we show the power spectra for species $x$ and $y$ respectively, for $N=50000$ and $\beta=2.92$. The value of the complex phase of the eigenvalues is marked by a vertical dashed line. These figures also show the oscillations found in the time series obtained from the stochastic difference equation. In these two figures we see good agreement between theory and simulation. However, as for the continuous-time case, non-Gaussian effects lead to larger discrepancies as one approaches the bifurcation point at $\beta=3$.


\subsection*{Beyond the stable fixed point}
\label{sec:qp}

\begin{figure}
\subfigure[]{\includegraphics[scale=0.9]{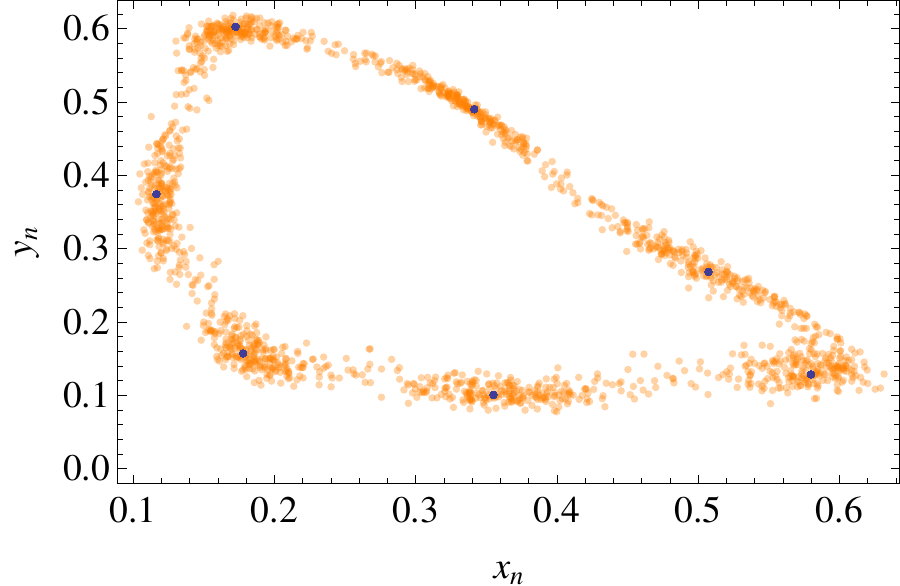}}
\subfigure[]{\includegraphics[scale=0.9]{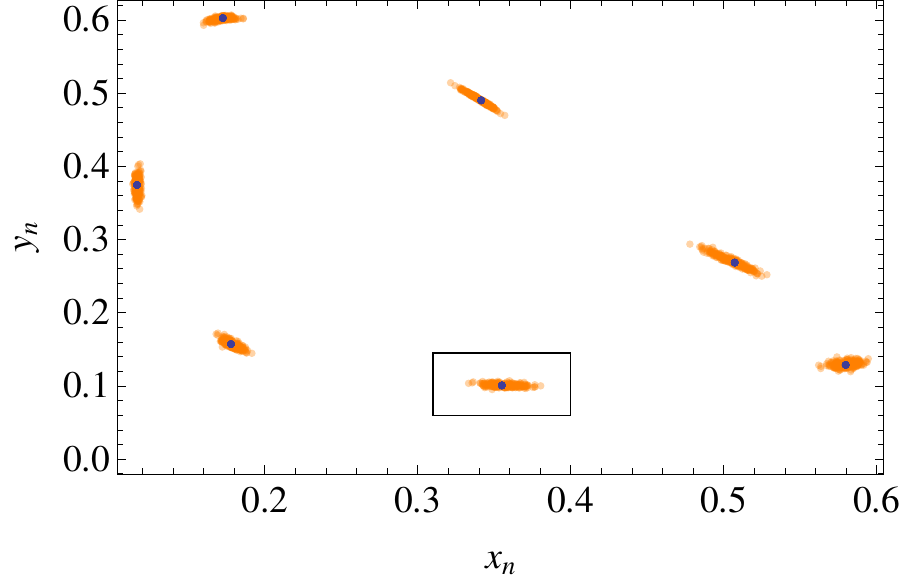}}\\
\subfigure[]{\includegraphics[scale=0.9]{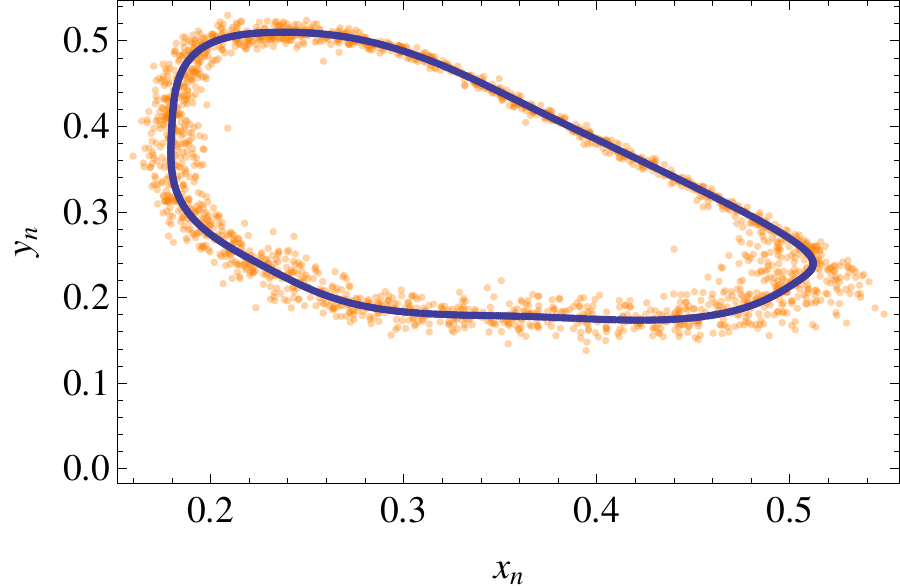}}
\subfigure[]{\includegraphics[scale=0.9]{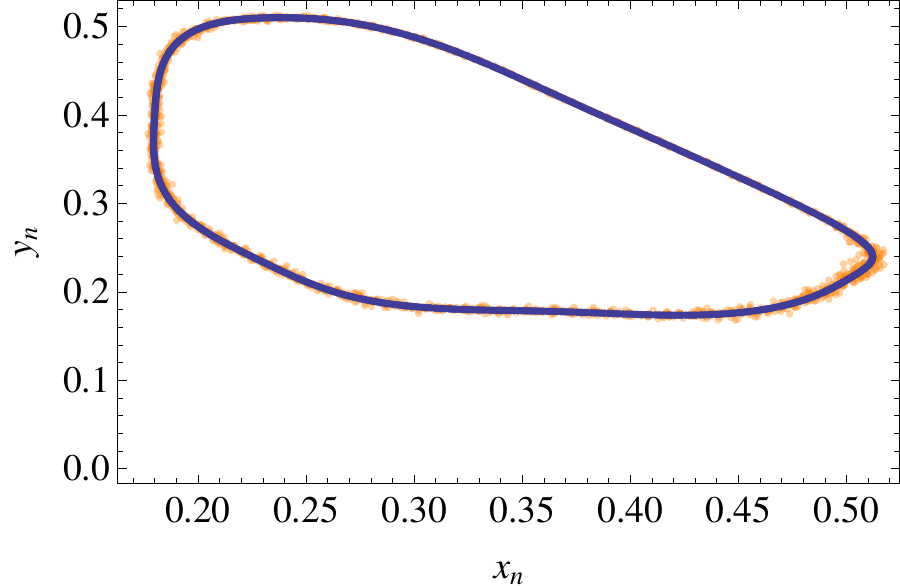}}
\caption{Comparison between the stochastic (orange) and deterministic (blue) dynamics for the Lotka-Volterra map. Top: the 7-cycle behavior ($\beta=3.6$) for (a) $N=10000$ and (b) $N=250000$. The boxed point in the right panel is further examined in Figure~\ref{fig:7_cycle_lin}, where the fluctuations around the point are quantified using the linear theory. Bottom: the fluctuations around the quasiperiodic attractor for $\beta=3.3$ for (c) $N=10000$ and (d) $N=250000$. In all cases $\mu=3$.}
\label{fig:qp_and_7}
\end{figure}

\begin{figure}[h!]
\subfigure[]{\includegraphics[scale=0.9]{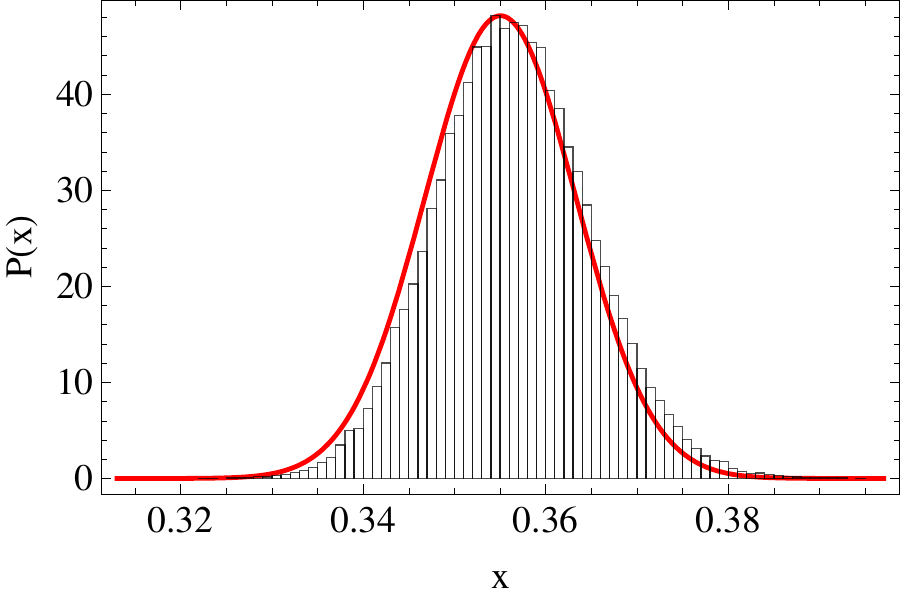}}
\subfigure[]{\includegraphics[scale=0.9]{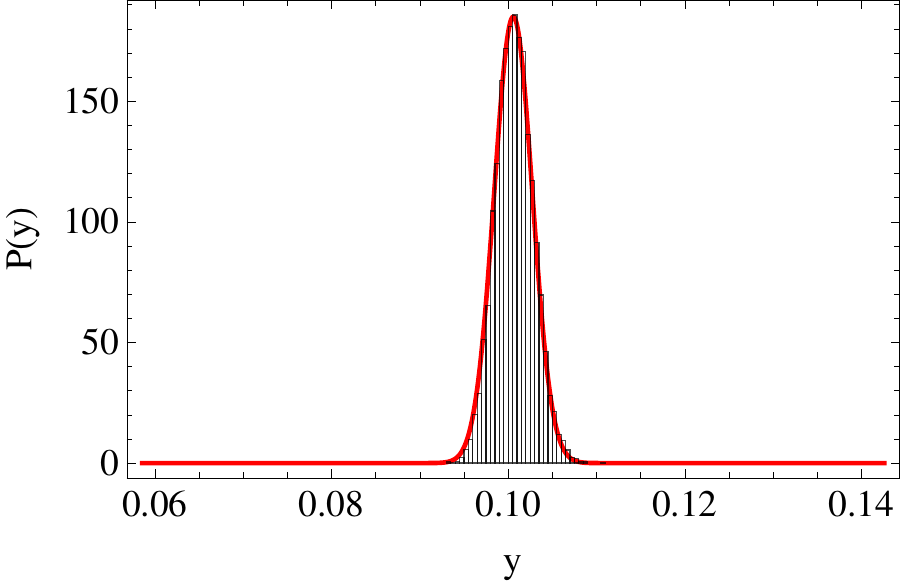}}\\
\caption{Fluctuations around one point of the 7-cycle, indicated by the box in Figure~\ref{fig:qp_and_7}(b). The parameter values are given in the caption of that figure. We have used the same scaling on the horizontal axis in each panel, to indicate the elongated nature of the fluctuations.}
\label{fig:7_cycle_lin}
\end{figure}

The results presented so far, for the linearization and the power spectra, have concentrated on the fluctuations around the fixed point. In this section we will look at the remaining parts of the bifurcation diagram, where the variety of observed behavior is much wider than for 1D maps (see \emph{e.g.} Chapter 3 of \cite{kaneko_86} for a detailed discussion). The range of these include: periodic behavior, quasiperiodic behavior and chaos. The periodic and quasiperiodic behaviors are closely related. If one looks at the system's time-evolution in the $x$-$y$ plane, in both cases one sees that closed orbits are formed, indicating oscillatory motion. In the periodic case, the system will return to the same collection of points each time, whereas this will not happen in the quasiperiodic case. Both instances can be found in the Lotka-Volterra map. The deterministic behavior is shown by the blue symbols in Figure~\ref{fig:qp_and_7}, with the top panels showing periodic behavior (in this case, a 7-cycle) and the lower panels the quasiperiodic motion. The orange symbols in the figure show a stochastic trajectory for a case of  strong (left panels) and weak (right panels) fluctuations. When the noise is weak, we can use our linear theory to find the stationary distribution for the fluctuations around the periodic attractor. Here we do this for one point of the 7-cycle, indicated by the box in Figure~\ref{fig:qp_and_7}(b). The distributions of the fluctuations for both species is shown in Figure~\ref{fig:7_cycle_lin}. Similar results may be found for the other 6 points that make up the attractor.


\begin{figure}
\subfigure[]{\includegraphics[scale=0.9]{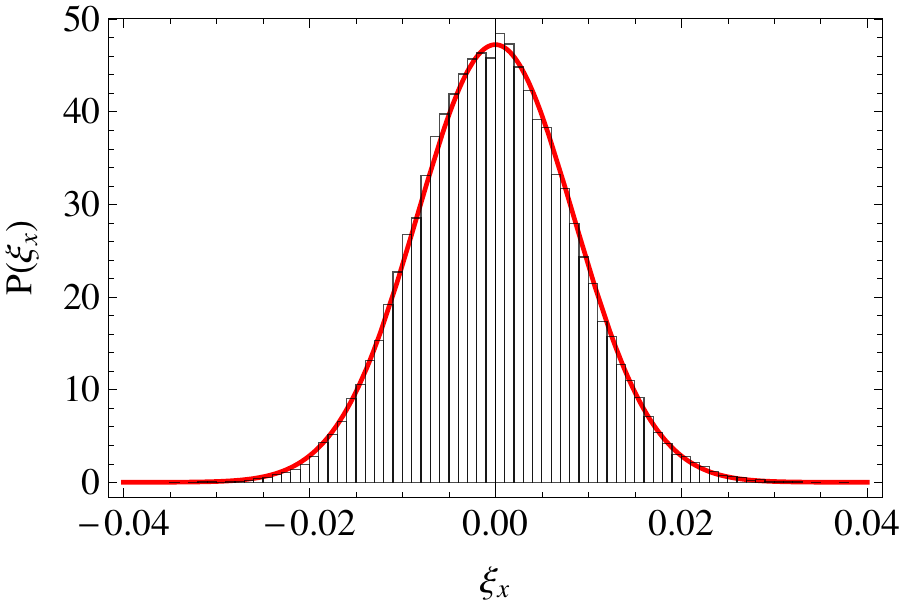}}
\subfigure[]{\includegraphics[scale=0.9]{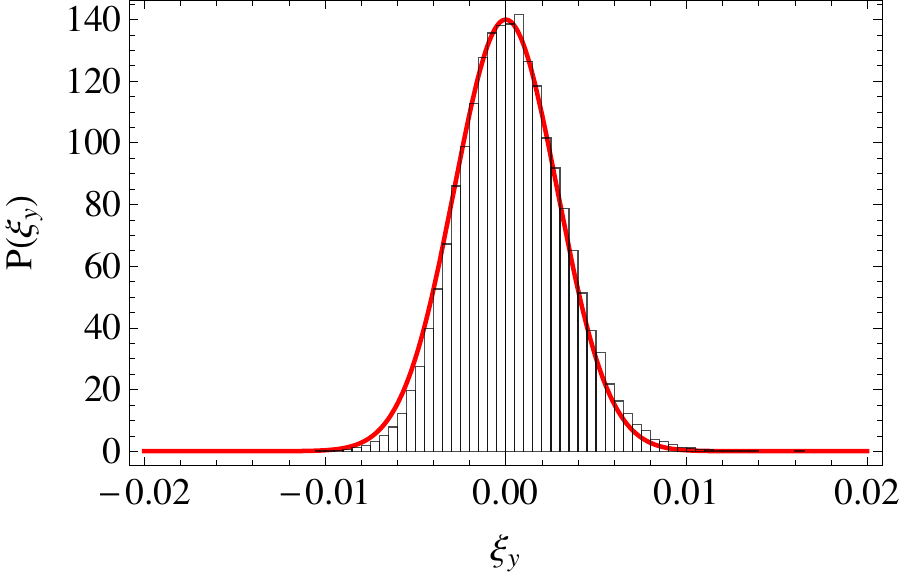}}\\
\caption{Fluctuations around the quasiperiodic attractor at $t=50$ for (a) $\xi_x$ and (b) $\xi_y$, comparing simulation data (bars) with the theoretical prediction (red line). The initial condition was placed on the attractor i.e. $\bm{\xi}_0=0$. Parameters used were $N=250000$, $\beta=3.3$ and $\mu=3$. For the simulation data, $8\times 10^4$ realizations of the process were generated, selecting one data point (at the desired time) from each.}
\label{fig:qp_lin}
\end{figure}


In the quasiperiodic case, in contrast with the case of a fixed point or periodic attractor, the variances of the fluctuations will grow with time, as in indicated in Figure~\ref{fig:qp_lin}. This is because there is a neutral direction (along the attractor), along which the fluctuations can diffuse. In continuous-time models, similar behavior can be seen in limit cycles \cite{tomita_74,boland_09}. So, from a given initial condition, we can estimate the distribution of fluctuations at a given time, using the linear theory. Distributions for both species are displayed in Figure~\ref{fig:qp_lin}, where they are compared with the simulation data. Here we write $(\xi_x,\xi_y)$ to denote the components of the fluctuations in the $x$- and $y$-directions respectively. To obtain the data, we simulate the stochastic difference equation, storing data at, or very close to the desired time, $t$. We can then build the distribution by averaging over an ensemble of trajectories.

\section{Chaos}
\label{sec:chaos}

\begin{figure}
\subfigure[]{\includegraphics[scale=0.9]{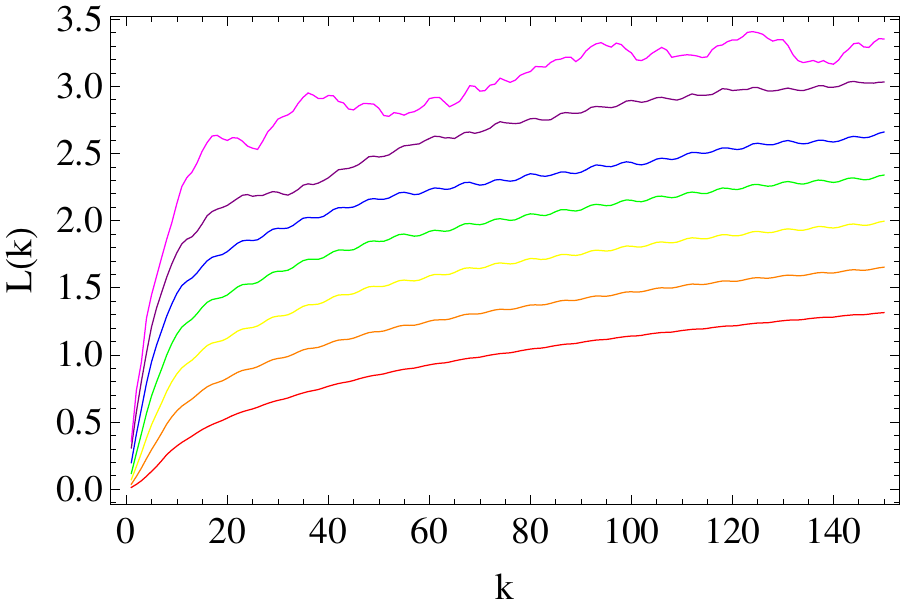}}
\subfigure[]{\includegraphics[scale=0.9]{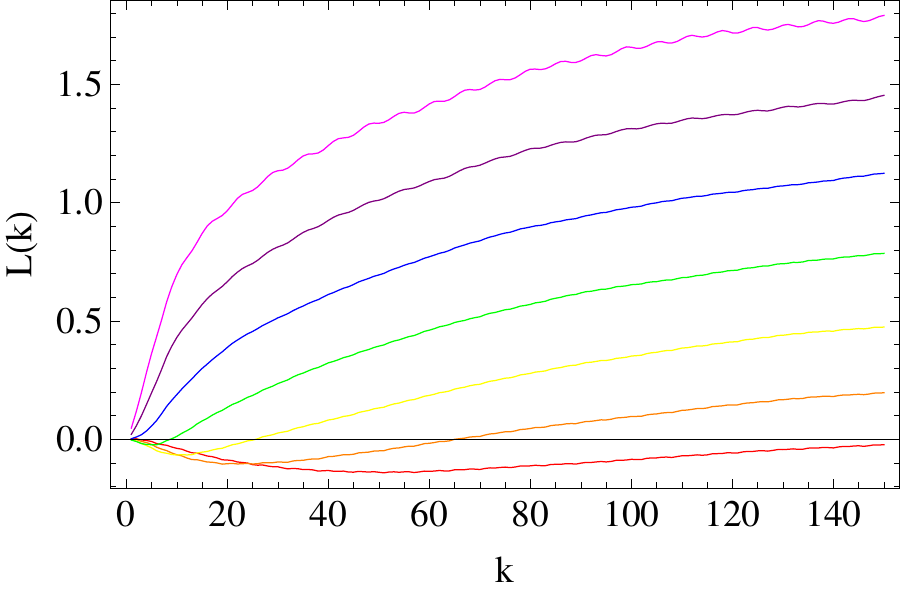}}\\
\caption{Time-dependent exponent curves from simulation data taken from the quasiperiodic regime for (a) $N=10000$ and (b) $N=250000$. Model parameters used were $\beta=3.3$ and $\mu=3$. The embedding parameters were $m=8$ and $l=1$. The curves were calculated using 7000 vectors $X_i$. The spatial scales used were $(2^{-(i+1)/2},2^{-i/2})$ with $i=10,11,\hdots,16$ (from bottom to top). In neither case do the curves collapse together to form an envelope. Therefore, the intrinsic stochasticity has not induced chaos-like dynamics.}
\label{fig:NIC_qp}
\end{figure}

There are many papers in the literature which have examined the effects of stochasticity on chaotic systems (see e.g. Refs.~\cite{crutchfield_80,mayer_81,crutchfield_82,kautz_85,geysermans_93,gao_99}).  One interesting question to examine concerns `noise-induced chaos'. It has been found~\cite{crutchfield_80,crutchfield_82,kautz_85,gao_99} that noise can, for parameter choices where the deterministic map is non-chaotic, induce a transition to behavior which bears the hallmarks of chaos. However, to the best of our knowledge, this has only been carried out for the case of external additive noise, rather than noise which is intrinsic to the system. To begin with we will examine the dynamics simply by studying the time series produced from the stochastic equations. That is, what would we make of the time series, if we did not know where it came from? We shall analyze the time series using the time delay embedding technique, designed to reconstruct a system's attractor from a single component of the time series, a result formalized as Taken's theorem \cite{takens_81}. From a time series of one species in the 2D map we construct the embedding vectors $X_i$ such that $X_i=(x_i,x_{i+l},x_{i+2l},\hdots,x_{i+(m-1)l})$ where the parameters $l$, the delay time, and $m$, the embedding dimension, must be suitably chosen \cite{ott_93,gao_94a}. We follow the work of Gao et al. \cite{gao_99} and calculate the time-dependent exponent curves, defined as
\begin{equation}
L(k)=\left\langle \textrm{ln} \left( \frac{||X_{i+k}-X_{j+k} ||}{||X_i-X_j||} \right) \right\rangle,
\label{l_of_k}
\end{equation}
where the averaging is performed over all pairs $(X_i,X_j)$, for which $||X_i-X_j||$ is found to lie within a prescribed small shell, denoted by $(r,r+\delta r)$. Calculating $L(k)$ over a range of shells, allows us to examine the system's behavior over a range of scales. For the case of deterministic chaos, the curves will increase linearly, before flattening. During the linear growth stage, the curves from different spatial scales collapse together, forming an envelope. The formation of this envelope has been used as a direct test for chaos, with its slope returning the value of the largest Lyapunov exponent \cite{gao_94a,gao_94b}. Gao et al. \cite{gao_99} examined the case of noise-induced chaos, using this method. Using a 1D map with additive noise they showed that an envelope was formed for a parameter choice for which the deterministic map is non-chaotic. This envelope was also found for the 1D intrinsic noise case \cite{challenger_14}. Here we will study the 2D case: we are especially interested in the details of the noise-induced transition, if it can be found. In particular we will test if noise-induced transitions can be found from a quasiperiodic attractor, and or the periodic case. We will start with the quasiperiodic case as, by definition, the largest Lyapunov exponent of deterministic map will be equal to zero. So, we will investigate if a transition to chaos, indicated by a positive exponent, can be induced in this case. Carrying out this numerical investigation we were unable to find noise-induced transitions to chaos from a quasiperiodic attractor. Some typical results are shown for the Lotka-Volterra system in Figure~\ref{fig:NIC_qp}, for the strong and weak noise cases. The panels on the left- and right-hand side show the curves $L(k)$ for each case, respectively: in neither do the curves collapse together to form an envelope.
 

\begin{figure}
\begin{center}
\subfigure[]{\includegraphics[scale=0.85]{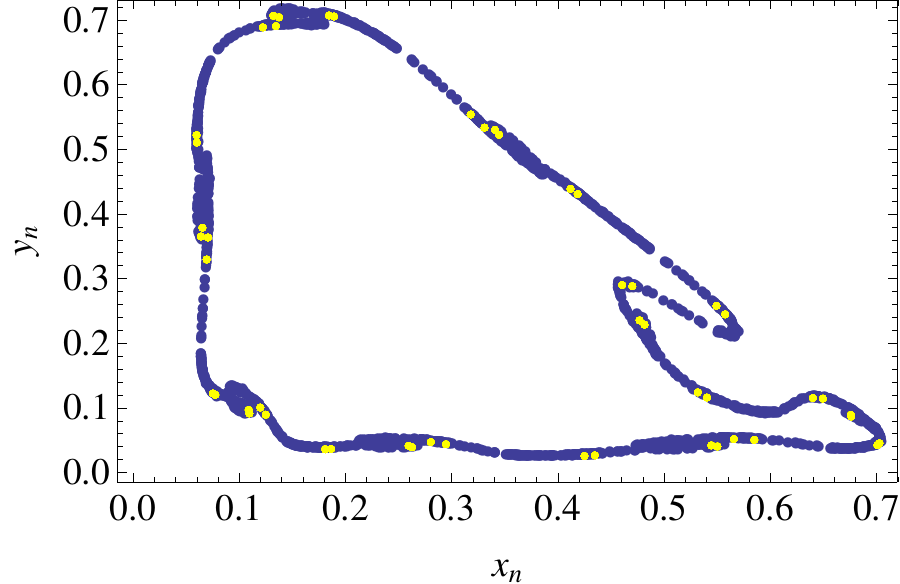}}
\subfigure[]{\includegraphics[scale=0.85]{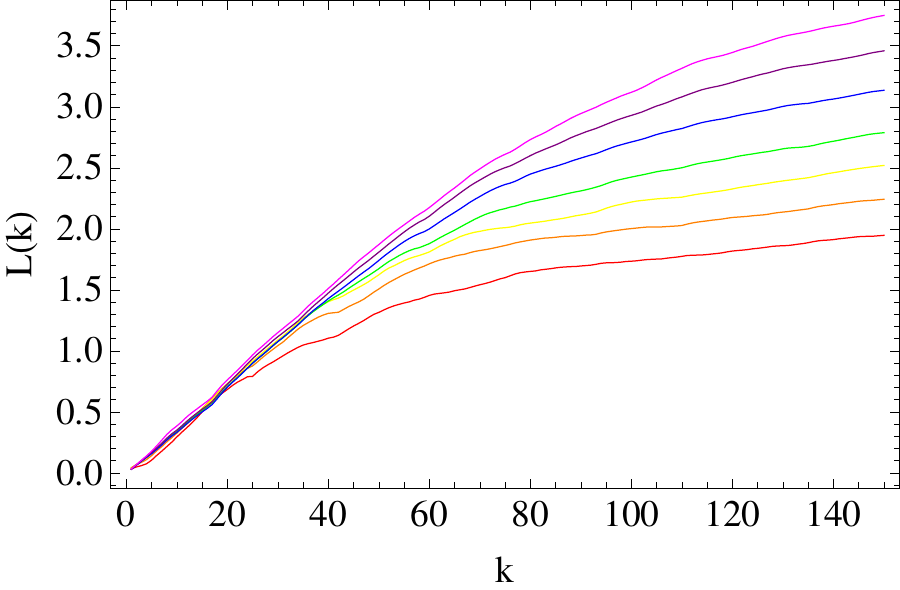}}
\end{center}
\caption{(a) Phase portrait comparing the stochastic (blue) and deterministic  (yellow) dynamics of the Lotka-Volterra map for the parameter choice $\beta=3.9116$, $\mu=3$. The system size used here was $N=10^7$. (b) Time-dependent exponent curves calculated from the stochastic time series. The spatial scales used were $(2^{-(i+1)/2},2^{-i/2})$ with $i=10,11,\hdots,16$ (from bottom to top). Embedding parameters were $m=8$ and $l=1$.}
\label{fig:NIC_LV}
\end{figure}

Our next step was to look for noise-induced chaos from periodic attractors. In Figure~\ref{fig:NIC_LV} we looked at the Lotka-Volterra map with parameters $\mu=3$ and $\beta=3.9116$. For these parameters, the deterministic map behaves periodically with a period of 50. The left-hand panel of the figure shows the deterministic dynamics in yellow, with a typical stochastic trajectory in blue, for $N=10^7$. The right-hand panel shows the curves $L(k)$ plotted over a range of spatial scales. The figure shows that, while the curves grow linearly (indicating exponential separation), the curves grow together, forming an envelope. This chaos-like behavior is visible over a range of $N$. However, if $N$ is extremely large (say, $10^9$), the stochastic dynamics instead follow the periodic attractor, and the envelope is not found. Or, if $N$ is too small, the stochastic effects dominate the dynamics and the `fingerprint' of chaotic behavior cannot be detected.

\begin{figure}
\begin{center}
\includegraphics[scale=1]{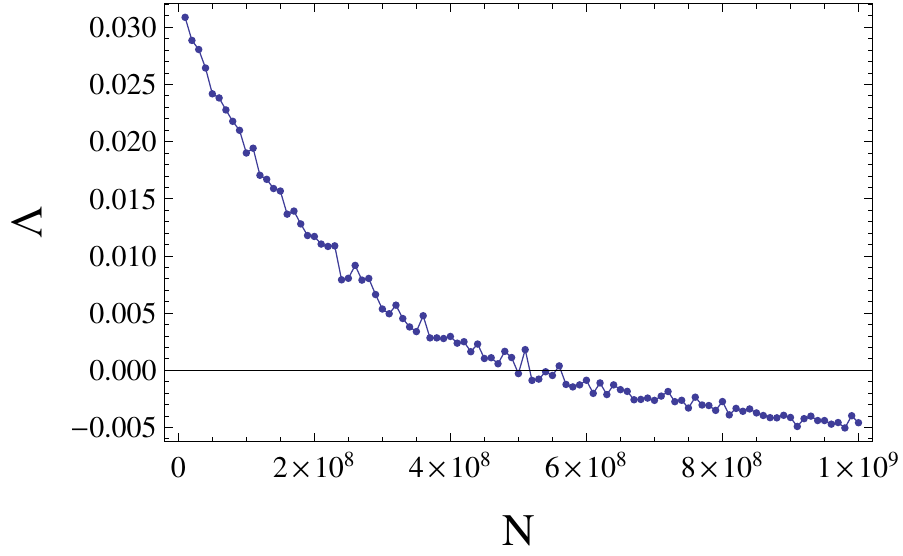}
\end{center}
\caption{Largest Lyapunov exponent as a function of $N$ for $\beta=3.9116$, $\mu=3$, calculated using the same formula as in Fig.~\ref{fig:lyap_vs_beta}.}
\label{fig:lyap_vs_N}
\end{figure}

To quantify the effect of the intrinsic noise one can alternatively apply the method employed in Ref.~\cite{crutchfield_82}, generalised to two dimensions (see e.g. \cite{ott_93}). This method uses the formula for the calculation of the Lyapunov exponents for a deterministic map. Hence, it assumes knowledge of the underlying dynamical process, in contrast with the method employed earlier. In Figure~\ref{fig:lyap_vs_N} we plot the largest Lyapunov exponent, $\Lambda$, as a function of $N$ for the same choice of parameters as Figure~\ref{fig:NIC_LV}. The noise-induced transition to chaos is clearly seen for a range of $N$ which agrees with that predicted by the method due to Gao et al.


\section{Discussion}
\label{sec:disc}

In this article we have extended previous work~\cite{challenger_13,challenger_14} to enable the effects of intrinsic noise in discrete-time systems to be studied in more than one dimension. We began with a microscopic model, and derived a mesoscopic description valid for large $N$, the maximum population that could be supported in the system. A key goal of the mesoscopic theory was finding the stochastic difference equation which describes the finite-$N$ system. The noise in this equation was intrinsic, and its form could only be found by starting from a microscopic description of the process. In the non-chaotic setting, we showed how linearization of this equation could provide analytical results for \emph{e.g.}~ the fluctuations around a fixed point. This linearized difference equation also provided the starting point for the calculation of the power spectra, which quantifies the sustained oscillations, driven by the stochasticity, which are visible around the fixed point of the deterministic map. In recent years, there has been much interest in this type of oscillation, often termed `quasi-cycles', in continuous-time systems. However, we are not aware of any work (either numerical or theoretical) which has indicated their presence in discrete-time models. This finding is interesting from a modeler's perspective, since if a real-world system is known to exhibit oscillations, this can inform the way in which the system is modeled (see Ref.~\cite{kendall_99} for an ecology-based discussion). 

Although we developed the theoretical framework for arbitrary dimensions, $d$, we restricted the applications of this theory to the $d=2$ case. Specifically, we investigated a 2D map with a Lotka-Volterra form, which we used to demonstrate the utility of the theory in a number of situations. The range of attractors found in 2D maps has much greater variety than their 1D counterparts, and so in addition to looking at linear fluctuations about the fixed point, we also investigated fluctuations about periodic and quasiperiodic attractors. In addition, we examined the ability of intrinsic noise to induce chaotic transitions from periodic and quasiperiodic attractors of the deterministic map. We found that periodic states close enough to parameter values for which the deterministic map behaves chaotically can present the hallmarks of chaos in the presence of intrinsic noise for a range of noise intensities; whereas we were unable to find such transition from quasiperiodic states.

The reason why the transition should be observed in one case (for the periodic attractor), but not in the other (the quasiperiodic attractor) is not clear and requires further study. In studies of this type \cite{crutchfield_82,gao_99} it is often remarked that noise-induced transitions to chaos are associated with parameter choices which are `close' to parameters for which the system is chaotic. In 2D maps, it is generally true \cite{kaneko_86} that the transitions to chaos come from period-doublings of periodic attractors, rather than directly from quasiperiodic states. Therefore it is logical that there will be periodic orbits which are very close to chaotic states. Using the right-hand panel of Figure~\ref{fig:NIC_LV} we can estimate the value of the exponent in this case, by calculating the slope ($\simeq$0.033) of the linear growth stage. This value is similar to the value of the exponent of nearby chaotic states. This can be seen by examining Figure~\ref{fig:lyap_vs_beta}, which shows the largest Lyapunov exponent for the deterministic map.

The work reported here suggests many possibilities for further investigation. From a theoretical viewpoint, it could be interesting to examine the effects of intrinsic noise on other, more exotic attractors found in discrete-time maps with $d>1$. These include the appearances of a number of chaotic `islands', or the coexistence of multiple quasiperiodic rings, which are visited sequentially (see Chapter 3 of Ref.~{\cite{kaneko_86} for a detailed study). From a practical point of view, it is now possible to connect our formalism with a wider range of ecological models, the majority of which contain multiple species. 

\begin{acknowledgments}
We wish to thank Antonio Politi for useful discussions. CP-R was funded by CONICYT Becas Chile scholarship No.~72140425. JDC and DF acknowledge support from the program Prin2012 financed by the Italian MIUR.
\end{acknowledgments}

\appendix
\section{Derivation of the discrete-time Fokker-Planck-like equation}
\label{sec:2-d-dim}

In this Appendix we will derive the analog of the Fokker-Planck equation which is found in Markov processes where the time is discrete. This is not completely straightforward because jumps from the variable $\mathbf{z}_t$ to $\mathbf{z}_{t+1}$ are not necessarily small, and so instead jumps from $\mathbf{p}_t$ to $\mathbf{z}_{t+1}$ need to be considered, where the function $\mathbf{p}$ defines the model, as explained in Section \ref{sec:theory}. The derivation closely follows that for the case of a single variable~\cite{challenger_14} and the reader is advised to initially consult this simpler derivation before reading the generalization given below to two variables, and then eventually to $d$ variables.

Beginning with the case of two variables, the starting point of the derivation is the Taylor expansion of the Chapman-Kolmogorov equation, given in Eq.~(\ref{CK}) where $\mathbf{z}=(z_1,z_2)$. Following a standard procedure \cite{gardiner_09}, we make the following change of variables,
\begin{eqnarray}
\label{CK1}
z_1' &=& z_1- \Delta z_1 \nonumber\\
z_2' &=& z_2- \Delta z_2,
\end{eqnarray}
so that the integrand of Eq.~\eqref{CK} can be written as:
\begin{equation}
\label{A}
P([z_1-\Delta z_1]+\Delta z_1,[z_2-\Delta z_2]+\Delta z_2,t+1 \mid 
 z_{1} - \Delta z_1,z_{2}-\Delta z_2,t) P( z_{1} - \Delta z_1,z_{2}-\Delta z_2,t \mid z_{10},z_{20},t_0).
\end{equation}
Performing a Taylor expansion of the above expression yields:
\begin{equation}
\label{A_taylor}
\sum_{\ell_1 = 0}^{\infty}\,\sum_{\ell_2 = 0}^{\infty} \frac{\left(-\Delta z_1 \right)^{\ell_1}}{\ell_{1}!}\,\frac{\left( -\Delta z_2 \right)^{\ell_2}}{\ell_{2}!}\,\frac{\partial^{\ell_1 + \ell_2}}{\partial^{\ell_1}z_1 \partial^{\ell_2} z_2 }
\left[ P(z_1+\Delta z_1,z_2+\Delta z_2,t+1 \mid z_{1},z_{2},t) P( z_{1},z_{2},t \mid z_{10},z_{20},t_0) \right].
\end{equation}
Let us now insert the above expression into the Chapman-Kolmogorov equation (\ref{CK}) 
\begin{eqnarray}
\label{CK_taylor}
& & P(\mathbf{z},t+1 \mid \mathbf{z_0},t_0) = \sum_{\ell_1 = 0}^{\infty}\,\sum_{\ell_2 = 0}^{\infty} \frac{(-1)^{\ell_1}}{\ell_{1}!}\,\frac{(-1)^{\ell_2}}{\ell_{2}!}\,\frac{\partial^{\ell_1 + \ell_2}}{\partial^{\ell_1}z_1 \partial^{\ell_2} z_2 } \times \\ \nonumber
& & \int \hspace{-0.2cm} \int d \left(\Delta z_1 \right)  d \left(\Delta z_2 \right)\left(\Delta z_1 \right)^{\ell_1} \left(\Delta z_2 \right)^{\ell_2} P(z_1+\Delta z_1,z_2+\Delta z_2,t+1 \mid z_{1},z_{2},t)\,P(\mathbf{z},t \mid \mathbf{z_0},t_0),
\end{eqnarray}
where $\mathbf{z_0}=(z_{10},z_{20})$. We now introduce the jump moments, which are defined as
\begin{equation}
\label{M}
M_{\ell_{1},\ell_{2}}(z_1,z_2) = \int \hspace{-0.2cm} \int  d w_1  d w_2  \left(w_1-z_1 \right)^{\ell_1} \left(w_2- z_2 \right)^{\ell_2} P(w_1,w_2,t+1 \mid z_{1},z_{2},t).
\end{equation}
Working in terms of these quantities, we arrive at 
\begin{equation}
\label{CK_taylor1}
P(\mathbf{z},t+1 \mid \mathbf{z_0},t_0) = \sum_{\ell_1 = 0}^{\infty}\,\sum_{\ell_2 = 0}^{\infty} \frac{(-1)^{\ell_1}}{\ell_{1}!}\,\frac{(-1)^{\ell_2}}{\ell_{2}!}\,\frac{\partial^{\ell_1 + \ell_2}}{\partial^{\ell_1}z_1 \partial^{\ell_2} z_2 } \left[ M_{\ell_1,\ell_2}(z_1,z_2)\,
P(\mathbf{z},t \mid \mathbf{z_0},t_0) \right].
\end{equation}

At this point we deviate from the standard derivation of the Fokker-Planck equation \cite{gardiner_09} and introduce jump moments from $(p_{1,t},p_{2,t})$ to $(z_{1,t+1},z_{2,t+1})$. To this end we first rewrite the jump moments
$M_{\ell_{1},\ell_{2}}(z_1,z_2)$ as a conditional average over the stochastic process $\mathbf{z}_t$:
\begin{equation}
\label{M2}
M_{\ell_1,\ell_2}(z_1,z_2) = \left\langle \left(z_{1,t+1}-z_{1,t}\right)^{\ell_1} \left(z_{2,t+1}-z_{2,t}\right)^{\ell_2}  \right\rangle_{\mathbf{z}_t = \mathbf{z}}.
\end{equation}
However we wish to recast the expansion in terms of the jump moments $J_{r_1,r_2}$ defined by Eq.~\eqref{J}, therefore we write Eq.~(\ref{M2}) in the form
\begin{eqnarray}
\label{M3}
M_{\ell_1,\ell_2}(z_1,z_2) &=& \left\langle \left[ z_{1,t+1}-p_{1,t}+(p_{1,t}-z_{1,t})\right]^{\ell_1} \left[z_{2,t+1}-p_{2,t}+(p_{2,t}-z_{2,t})\right]^{\ell_2}  \right\rangle_{\mathbf{z}_t = \mathbf{z}}, \nonumber \\
&=& \sum_{r_1 = 0}^{\ell_1}\,\sum_{r_2 = 0}^{\ell_2}\,\binom{\ell_1}{r_1}\,\binom{\ell_2}{r_2}\, \left( p_1-z_1 \right)^{\ell_1 - r_1}\,\left( p_2-z_2 \right)^{\ell_2 - r_2}\,J_{r_1, r_2}(p_1,p_2),
\end{eqnarray}
where 
\begin{equation}
\label{M4}
J_{r_1,r_2}(p_1,p_2) = \left\langle \left(z_{1,t+1}-p_{1}\right)^{r_1} \left(z_{2,t+1}-p_{2}\right)^{r_2} \right\rangle_{\mathbf{z}_t = \mathbf{z}},
\end{equation}
as given in Eq.~\eqref{J}, and where $\mathbf{p} \equiv \left. \mathbf{p}_t\right|_{\mathbf{z}_t = \mathbf{z}}$.
 
We can calculate the jump moments $J_{r_1,r_2}$ from the formulation of the model in terms of a Markov chain. In this case, the maximum population size, $N$, is finite, and the states are defined by $n_{i,t} = N z_{i,t}$, where $i=1,2$ and $n_{i,t}$ is the number of individuals of type $i$ at time $t$. Therefore we may write Eq.~\eqref{M4} as
\begin{equation}
\label{J_n}
J_{r_1,r_2}(p_1,p_2) = \frac{1}{N^{r_1}} \frac{1}{N^{r_2}}\,\left\langle \left[ n_{1,t+1}-\left( N p_{1}\right) \right]^{r_1} \left(n_{2,t+1}- \left( Np_{2}\right) \right]^{r_2}  \right\rangle_{\mathbf{n}_t = \mathbf{m}}.
\end{equation}
Now the conditional probability that the system is in state $\mathbf{n}$ at
time $t+1$, given it was in state $\mathbf{m}$ at time $t$ is simply the transition matrix $Q_{\mathbf{n};\mathbf{m}}$, and so we may write Eq.~\eqref{J_n} as
\begin{equation}
\label{J_n_exp}
J_{r_1,r_2}(p_1,p_2) = \frac{1}{N^{r_1}} \frac{1}{N^{r_2}}\,\sum_{n_1}\,
\sum_{n_2}\,\left[ n_{1}-\left( N p_{1}\right) \right]^{r_1} \left[n_{2}- \left( Np_{2}\right) \right]^{r_2}\,Q_{n_1,n_2;m_1,m_2}.
\end{equation}
(Note that the corresponding equation in \cite{challenger_14} [Eq.~(52)] has a factor of $(-1)^r$ omitted). Since $Q_{n_1,n_2;m_1,m_2}$ is just the trinomial distribution \eqref{trinomial}, the $J_{r_1,r_2}$ are just the central moments~\cite{grimmett_92} of the trinomial distribution. We show in Appendix \ref{sec:j_d} that $J_{r_1,r_2} = \mathcal{O}(N^{-2})$ if $r_1 + r_2 > 2$, and so these terms will, as usual, be neglected when setting up the Fokker-Planck-like equation. Of the remaining $J_{r_1,r_2}$, $J_{0,0}=1$ by normalization; $J_{1,0}$ and $J_{2,0}$ can be found by performing the $n_2$ sum and so finding the first two central moments of a binomial distribution, giving $J_{1,0}=0$ and $J_{2,0} = N^{-1} p_1(1-p_1)$; $J_{0,1}$ and $J_{0,2}$ can be similarly found to be given by $J_{0,1}=0$ and $J_{0,2} = N^{-1} p_2(1-p_2)$. Only the calculation of $J_{1,1}$ requires explicit use of the trinomial distribution, giving $J_{1,1} = - N^{-1} p_1 p_2$. Substituting these value
 s for $J_{r_1,r_2}$
  into Eq.~\eqref{M3} gives
\begin{eqnarray}
\label{M5}
& & M_{\ell_1, \ell_2}(z_1,z_2) =  \left(p_1-z_1\right)^{\ell_1} \left(p_2-z_2\right)^{\ell_2} + \frac{\ell_1(1-\ell_1)}{2}\,N^{-1} p_1\left( 1 - p_1 \right)\,\left(p_1-z_1\right)^{\ell_1 -2} \left(p_2-z_2\right)^{\ell_2} \\ \nonumber
& & + \frac{\ell_2(1-\ell_2)}{2}\,N^{-1} p_2\left( 1 - p_2 \right)\,\left(p_1-z_1\right)^{\ell_1} \left(p_2-z_2\right)^{\ell_2 - 2}  - \ell_1 \ell_2\,N^{-1} p_1 p_2
\left(p_1-z_1\right)^{\ell_1 -1} \left(p_2-z_2\right)^{\ell_2 - 1} + 
\mathcal{O}(N^{-2}).
\end{eqnarray}

We now substitute the expression \eqref{M5} into \eqref{CK_taylor1} to find
an expression for $P(\mathbf{z},t+1 \mid \mathbf{z_0},t_0)$ in terms of $P(\mathbf{z},t \mid \mathbf{z_0},t_0)$. There are four terms, corresponding to the 
four terms in Eq.~\eqref{M5} which, dropping the dependence of these two pdfs on the initial conditions and writing $P_{t}(\mathbf{z})$ for $P(\mathbf{z},t)$, we may write as $P_{t+1}(\mathbf{z}) =  P^{(1)}_{t+1}(\mathbf{z}) + P^{(2)}_{t+1}(\mathbf{z})+  P^{(3)}_{t+1}(\mathbf{z})+ P^{(4)}_{t+1}(\mathbf{z})$. For illustrative purposes, let us write down explicitly the second of these terms:
\begin{eqnarray}
\label{CK_fin}
P^{(2)}_{t+1}(\mathbf{z}) &=& \sum_{\ell_1 = 0}^{\infty}\,\sum_{\ell_2 = 0}^{\infty} \frac{(-1)^{\ell_1}}{\ell_{1}!}\,\frac{(-1)^{\ell_2}}{\ell_{2}!}\,\frac{\ell_1\left(\ell_1 - 1 \right)}{2N}\,p_1\left( 1 - p_1 \right)\,\frac{\partial^{\ell_1 + \ell_2}}{\partial^{\ell_1}z_1 \partial^{\ell_2} z_2 } \left(p_1-z_1\right)^{\ell_1 -2} \left(p_2-z_2\right)^{\ell_2} P_t(\mathbf{z}) \nonumber \\
&=& \frac{1}{2N}\,\sum_{\ell^{'}_1 = 0}^{\infty}\,\sum_{\ell_2 = 0}^{\infty} \frac{(-1)^{\ell^{'}_{1}}}{\ell^{'}_{1}!}\,\frac{(-1)^{\ell_2}}{\ell_{2}!}\, p_1\left( 1 - p_1 \right)\,\frac{\partial^{\ell^{'}_1 + \ell_2 + 2}}{\partial^{\ell^{'}_1 + 2}z_1 \partial^{\ell_2} z_2 } \left(p_1-z_1\right)^{\ell^{'}_1} \left(p_2-z_2\right)^{\ell_2} P_t(\mathbf{z}).
\end{eqnarray}

As in the case of one variable~\cite{challenger_14}, this expression can be simplified by first taking its Fourier transform:
\begin{eqnarray}
\label{Fourier1}
\nonumber
&&\tilde{P}^{(2)}_{t+1}(\mathbf{k}) = \frac{1}{2N}\,\sum_{\ell^{'}_1 = 0}^{\infty}\,\sum_{\ell_2 = 0}^{\infty} \frac{(-1)^{\ell^{'}_{1}}}{\ell^{'}_{1}!}\,\frac{(-1)^{\ell_2}}{\ell_{2}!}\,\int_{-\infty}^{\infty}\int_{-\infty}^{\infty} d  \mathbf{z} \exp(i \mathbf{k} \cdot \mathbf{z})\,\frac{\partial^{\ell^{'}_1 + \ell_2 + 2}}{\partial^{\ell^{'}_1 + 2}z_1 \partial^{\ell_2} z_2 } \left(p_1-z_1\right)^{\ell^{'}_1} \left(p_2-z_2\right)^{\ell_2} p_1\left( 1 - p_1 \right) P_t(\mathbf{z}) \\ \nonumber
&=& \frac{1}{2N}\,\sum_{\ell^{'}_1 = 0}^{\infty}\,\sum_{\ell_2 = 0}^{\infty} \frac{1}{\ell^{'}_{1}!}\,\frac{1}{\ell_{2}!}\,\int_{-\infty}^{\infty}\int_{-\infty}^{\infty} d  \mathbf{z} \exp(i \mathbf{k} \cdot \mathbf{z})\,\left( ik_1 \right)^{2}\,\left[ \left(p_1-z_1\right) \left( ik_1 \right) \right]^{\ell^{'}_1} \left[ \left(p_2-z_2\right) \left( i k_2 \right) \right]^{\ell_2}\,p_1\left( 1 - p_1 \right) P_t(\mathbf{z}) 
\\ \nonumber
&=& \frac{1}{2 N}\,\int_{-\infty}^{\infty}\int_{-\infty}^{\infty} d  \mathbf{z} \exp(i \mathbf{k} \cdot \mathbf{z})\exp(i \mathbf{k} \cdot (\mathbf{p}-\mathbf{z}))\,\left( i k_1 \right)^{2}\,p_1(1-p_1)  P_t(\mathbf{z}) \\ 
&=& \frac{1}{2 N}\,\int_{-\infty}^{\infty}\int_{-\infty}^{\infty} d  \mathbf{z} \exp(i \mathbf{k} \cdot \mathbf{p})\,\left( i k_1 \right)^{2}\,p_1(1-p_1) P_t(\mathbf{z}).
\end{eqnarray}
Since the only $\mathbf{z}$ dependence (other than through $\mathbf{p}$) in the integrand of Eq.~(\ref{Fourier1}) is in $P_t(\mathbf{z})$, we may write $P_t(\mathbf{z})\,d\mathbf{z}$ as $\mathcal{P}_t(\mathbf{p})\,d\mathbf{p}$, such that $P_t(\bm{z})=\mathcal{P}_t(\bm{p})|\textrm{det}(\bm{J})|$ and $J_{ij}=\partial p_i/\partial z_j$. Therefore
\begin{eqnarray}
\label{Fourier2}
\tilde{P}^{(2)}_{t+1}(\mathbf{k}) &=& \frac{1}{2 N}\,\int_{-\infty}^{\infty}\int_{-\infty}^{\infty} d  \mathbf{p} \exp(i \mathbf{k} \cdot \mathbf{p})\,\left( i k_1 \right)^{2}\,\left[ p_1(1-p_1) \mathcal{P}_t(\mathbf{p}) \right] \nonumber \\
&=& \frac{1}{2 N}\,\int_{-\infty}^{\infty}\int_{-\infty}^{\infty} d  \mathbf{p} \exp(i \mathbf{k} \cdot \mathbf{p})\frac{\partial^2}{\partial^2 p_1 }\left[ p_1(1-p_1) \mathcal{P}_t(\mathbf{p}) \right].
\end{eqnarray}
Taking the inverse Fourier transform gives the result
\begin{equation}
\label{inv_Fourier1}
P^{(2)}_{t+1}(\mathbf{z}) = \frac{1}{2 N}\,\frac{\partial^2}{\partial^2 z_1 }\left[ z_1(1-z_1) \mathcal{P}_t(\mathbf{z}) \right].
\end{equation}
The other terms can be treated in complete analogy, giving 
\begin{eqnarray}
\label{inv_Fourier2}
P_{t+1}(\mathbf{z}) = \left[ 1 + \frac{1}{2 N}\,\frac{\partial^2}{\partial^2 z_1 } z_1 \left( 1-z_1 \right) + \frac{1}{2 N}\,\frac{\partial^2}{\partial^2 z_2} z_2\left( 1 - z_2 \right) - \frac{1}{N}\,\frac{\partial^2}{\partial z_1 \partial z_2} z_1 z_2 + \mathcal{O}(N^{-2}) \right]\,\mathcal{P}_t(\mathbf{z}),
\end{eqnarray}
which is the required form for the evolution of the probability distribution, and may be written in the form given in Eq~\eqref{J_final}, after the usual neglect of the $\mathcal{O}(N^{-2})$ terms.

An equivalent formalism is to use stochastic difference equations. One begins by postulating that such equations should have the form $z_{i,t+1} = p_{i,t} +$\ noise, where the noise, $\eta_{i,t}$ is Gaussian with zero mean, and a correlation function $\langle \eta_{i,t}\,\eta_{i',t'} \rangle = N^{-1}\,B_{i i'}\,\delta_{t,t'}$. The function $B_{i i'}$ has to be chosen so that the stochastic process is equivalent to \eqref{inv_Fourier2}. The way to do this is to calculate the jump moments defined by \eqref{M4} using $z_{i,t+1} - p_{i,t} = \eta_{i,t}$. That is,
\begin{equation}
\label{M4_stoc}
J_{r_1,r_2}(p_1,p_2) = \left\langle \left(\eta_{1,t}\right)^{r_1} \left(\eta_{2,t} \right)^{r_2} \right\rangle_{\mathbf{z}_t = \mathbf{z}}.
\end{equation}
Clearly $J_{0,0} = 1$, and $J_{1,0}=J_{0,1}=0$ as required, $J_{2,0} = N^{-1}B_{1 1}, J_{0,2} = N^{-1}B_{2 2}$ and $J_{1,1} = N^{-1}B_{1 2} = N^{-1}B_{2 1}$. To obtain agreement with the jump moments calculated from the Markov chain we require that 
\begin{equation}
B_{i i} = p_i \left( 1 - p_i \right), \ \ B_{i j} = - p_i p_j, \ \ i,j=1,2, \ \ i \neq j.
\label{Bs}
\end{equation}
This is the form given in \eqref{sde_correlator} of the main text. In addition, we see from \eqref{M4_stoc} that $J_{r_1,r_2}$ is of order $N^{-2}$ for $r_1 + r_2 > 2$.

The generalization of this derivation from 2 variables to the general case of $d$ variables follows very similar lines, and we will simply indicate how the results at various stages of the derivation differ. We will frequently adopt a vector notation, for instance, $\bm{z}=(z_1,z_2,\hdots,z_d), \bm{\ell} = (\ell_1,\ell_2,\hdots,\ell_d), \mathbf{r} = (r_1,r_2,\hdots,r_d)$, etc.

Again, our starting point is Eq.~\eqref{CK}, but for $\bm{z}=(z_1,z_2,\hdots,z_d)$. We introduce $\bm{\Delta z}=\bm{z}-\bm{z}'$ and make a Taylor expansion of the integrand:
\begin{equation}
\label{A_taylor_d}
\left( \prod^{d}_{i=1} \sum_{\ell_i = 0}^{\infty}\,\frac{\left(-\Delta z_i \right)^{\ell_i}}{\ell_{i}!} \right)\,\frac{\partial^{\sum^{d}_{i=1} \ell_i}}{\partial^{\ell_1}z_1 \hdots \partial^{\ell_d} z_d }\,\left[ P(\bm{z}+\bm{\Delta z},t+1|\bm{z},t)P(\bm{z},t|\bm{z_0},t_0) \right].
\end{equation}
This can be rewritten as
\begin{equation}
\label{CK_taylor1_d}
P(\mathbf{z},t+1 \mid \mathbf{z_0},t_0) = \left( \prod^{d}_{i=1} \sum_{\ell_i = 0}^{\infty}\,\frac{\left( - 1 \right)^{\ell_i}}{\ell_{i}!} \right)\,\frac{\partial^{\sum^{d}_{i=1} \ell_i}}{\partial^{\ell_1}z_1 \hdots \partial^{\ell_d} z_d }\,\left[ M_{\bm{\ell}}(\bm{z}) P(\bm{z},t|\bm{z_0},t_0) \right],
\end{equation}
where we have introduced
\begin{eqnarray}
 M_{\bm{\ell}}(\bm{z}) &=& \int\hdots\int \left( \prod^{d}_{i=1} dw_{i} \left(w_i - z_i \right)^{\ell_i} \right)\,P(\bm{w},t+1|\bm{z},t) \nonumber \\
&=& \left\langle \left( \prod^{d}_{i=1} \left( z_{i,t+1} - z_{i,t} \right)^{\ell_i} \right) \right\rangle_{\mathbf{z}_t = \mathbf{z}}.
\end{eqnarray}
Just as in the case of one and two variables, these jump moments are not the appropriate ones; the jumps $z_{i,t+1} - z_{i,t}$ are not necessarily small, however $z_{i,t+1} - p_{i,t}$ are. We therefore write $z_{i,t+1} - z_{i,t} = z_{i,t+1} - p_{i,t} + (p_{i,t} - z_{i,t})$, and use the fact that the jump moments are conditional on $z_{i,t}$, (and so $p_{i,t}$) being given, and equal to say $z_i$ and $p_i$ respectively. Thus $z_{i,t+1} - p_{i,t}$ is equal to $z_{i,t+1} - z_{i,t}$ up to a shift $(p_{i} - z_{i})$. Expanding in terms of these shifts leads to
\begin{equation}
\label{M3_d}
 M_{\bm{\ell}}(\bm{z}) = \left( \prod^{d}_{i=1} \sum^{\ell_i}_{r_i = 0}\,\binom{\ell_i}{r_i}\,\left( p_i - z_i \right)^{\ell_i - r_i} \right) J_{\bm{r}}(\bm{p}), 
\end{equation}
where 
\begin{eqnarray}
J_{\bm{r}}(\bm{p}) &=& \left\langle \left( \prod^{d}_{i=1} \left(z_{i,t+1}-p_{i}\right)^{r_i} \right) \right\rangle_{\mathbf{z}_t = \mathbf{z}} \nonumber \\
&=& \left( \prod^{d}_{i=1} \frac{1}{N^{r_i}}\,\sum_{n_i} \left[ n_i - \left( N p_i \right) \right]^{r_i} \right)\,Q_{\bm{n};\bm{m}}.
\label{M4_d}
\end{eqnarray}

The distribution $Q_{\bm{n};\bm{m}}$ is a multinomial, and from \eqref{M4_d} $J_{\bm{r}}$ can be found in terms of its moments. In Appendix~\ref{sec:j_d} we will show that the $J_{\bm{r}}$ are of $\mathcal{O}\left(N^{-2}\right)$ for $\sum_{i=1}^d r_i > 2$, and so will not enter into the mesoscopic description that we are constructing. If $\sum_{i=1}^d r_i = 0$, all $r_i$ are zero, and the corresponding value of $J$ equals 1, by normalization. If $\sum_{i=1}^d r_i = 1$, all but one $r_i, r_j$ say, is non-zero, and $r_j = 1$. Then all the sums in \eqref{M4_d} apart from over $n_j$ can be carried out, leaving one to find the mean of a binomial variable, shifted by its mean, which gives zero. Finally, if $\sum_{i=1}^d r_i = 2$, one possibly is that all but one $r_i, r_j$ say, is non-zero, and $r_j = 2$. Then by the same argument, the problem reduces to finding the variance of a binomial distribution, giving $N^{-1}p_j(1-p_j)$. The other possibility is that all but two $r_i, r_j$ and $r_k$ say, are non-zero, and $r_j = r_k = 1$. Then by summing out all the variables but $n_j$ and $n_k$, the distribution is reduced to a trinomial, and one can find as in the case of two variables that the corresponding $J$ is equal to $-N^{-1}p_j p_k$. This leads to the analog of Eq.~(\ref{M5}):
\begin{eqnarray}
\label{M5_d}
M_{\bm{\ell}}(\bm{z}) &=& \prod^{d}_{i=1} \left(p_i-z_i\right)^{\ell_i} 
+ \sum^{d}_{j=1} \frac{\ell_j(1-\ell_j)}{2}\,N^{-1} p_j\left( 1 - p_j \right)\,\left(p_j-z_j\right)^{\ell_j -2}\,\prod^{d}_{i \neq j} \left(p_i-z_i\right)^{\ell_i} 
\\ \nonumber
&-& \sum^{d}_{j,k;\, j>k} \ell_j \ell_k \,N^{-1} p_j p_k\,\left(p_j-z_j\right)^{\ell_j -1}\,\left(p_k-z_k\right)^{\ell_k -1} \prod^{d}_{i \neq j,k} \left(p_i-z_i\right)^{\ell_i} + \mathcal{O}(N^{-2}).
\end{eqnarray}
As already mentioned, the $J_{\bm{r}}$ that contribute are exactly those that are found in the case of two variables, where $Q_{\bm{n};\bm{m}}$ is a trinomial distribution. As a consequence, \eqref{M5_d} has exactly the same structure as \eqref{M5} and so the steps from \eqref{CK_fin} to \eqref{inv_Fourier1} are essentially identical in the general case of $d$ variables. Hence we immediately arrive at the analog of \eqref{inv_Fourier2}
\begin{eqnarray}
\label{inv_Fourier2_d}
P_{t+1}(\mathbf{z}) = \left[ 1 + \frac{1}{2 N}\sum^{d}_{j=1}\,\frac{\partial^2}{\partial^2 z_j } z_j \left( 1-z_j \right) - \frac{1}{N}\sum^{d}_{j>k}\,\frac{\partial^2}{\partial z_j \partial z_k} z_j z_k + \mathcal{O}(N^{-2}) \right]\,\mathcal{P}_t(\mathbf{z}).
\end{eqnarray}
Neglecting the $\mathcal{O}(N^{-2})$ terms, and rewriting slightly gives
\begin{equation}
\label{J_final_gen}
P_{t+1}(\mathbf{z}) =  \mathcal{P}_t(\mathbf{z}) + \frac{1}{4 N } \sum_{i,j=1}^d \frac{\partial^2}{\partial z_i \partial z_j} \left[ \left(z_i(\delta_{ij}-z_j)+ z_j(\delta_{ij}-z_i)\right) \mathcal{P}_t(\mathbf{z}) \right].
\end{equation}
This is Eq.~\eqref{J_final} in the main text.

The equivalent formalism using stochastic difference equations is straightforward to determine. Writing $z_{i,t+1} - p_{i,t} = \eta_{i,t}$, as in the case of 2 variables, one finds from \eqref{M4_d} that
\begin{equation}
\label{M4_stoc_d}
J_{r_1,r_2}(p_1,p_2) = \left\langle \left( \prod^{d}_{i=1} \left(\eta_{i,t}\right)^{r_i} \right) \right\rangle_{\mathbf{z}_t = \mathbf{z}},
\end{equation}
which once again leads to 
\begin{equation}
B_{i i} = p_i \left( 1 - p_i \right), \ \ B_{i j} = - p_i p_j, \ \ i,j=1,\hdots,d, \ \ i \neq j,
\label{Bs_d}
\end{equation}
and $J_{\bm{r}}$ being of order $N^{-2}$ for $\sum^{d}_{i=1} r_i > 2$.

\section{Jump moments $J_{\mathbf{r}}\left(\mathbf{p} \right)$ for $\sum^{d}_{i=1} r_{i} > 2$.}
\label{sec:j_d}

In the derivation of the discrete-time analog of the Fokker-Planck equation in $d$ dimensions, carried out in Appendix~\ref{sec:2-d-dim}, we have used the fact that the moments given by $J_{\mathbf{r}}\left(\mathbf{p} \right)$ defined by \eqref{M4_d} are of $\mathcal{O}\left(N^{-2}\right)$ for $\sum_{i=1}^d r_i > 2$. Here, we prove this statement.

Expanding out \eqref{M4_d}, and performing the average gives
\begin{equation}
 J_{\mathbf{r}}\left(\mathbf{p} \right) = \prod^{d}_{i=1} \sum^{r_i}_{s_i = 0} \binom{r_i}{s_i}\,\left( - 1 \right)^{r_i - s_i}\,p_i^{r_i - s_i}\,\frac{\mu_{\bm{s}}}{N^S},
\label{eqn:Jr}
\end{equation}
where $S \equiv \sum^{d}_{i=1} s_i$ and where $\mu_{\mathbf{s}}$ is the moment $\left\langle n_1^{s_1}\cdots n_d^{s_d} \right\rangle$ of the multinomial distribution, with $n_i = N z_i$. 

To find the dependence of $\mu_{\bm{s}}$ on $N$, we first find the $N$ dependence of the factorial moments defined by~\cite{hogg_13}
\begin{equation}
\nu_{\mathbf{s}} = \left\langle n_1(n_1-1)\hdots (n_1-s_1+1)n_2(n_2-1)\cdots (n_2-s_2+1)\cdots n_d(n_d-1)\cdots (n_d-s_d+1)\right\rangle.
\label{fact_def}
\end{equation}
It is straightforward to prove that 
\begin{equation}
\nu_{\mathbf{s}} = N(N-1) \hdots (N-S+1)p^{s_1}_{1} \hdots p^{s_d}_{d},
\label{fact_result}
\end{equation}
as we now show. To do this we introduce the generating function for the multinomial distribution~\cite{hogg_13}
\begin{align}
\phi(\bm{p},\bm{w})=\sum_{n_1=0}^{N}\hdots \sum_{n_d=0}^{N}\frac{N!}{n_1!\hdots n_d!(N-n_1-\hdots -n_d)!}&p_1^{n_1}\hdots p_d^{n_d}(1-p_1-\hdots-p_d)^{N-n_1-\hdots -n_d}w_1^{n_1}\hdots w_d^{n_d} \nonumber \\
=&[p_1w_1+p_2w_2+\hdots+p_dw_d+(1-p_1-p_2\hdots p_d)]^N .\nonumber \\
\end{align}
Operating with $\partial^{S}/\partial w_1^{s_1}...\partial w_d^{s_d}$
on $\phi(\bm{p},\bm{w})$, and setting each entry in $\bm{w}$ to unity returns the desired result in Eq.~\eqref{fact_result}. 

Expanding the factorial moments in Eq.~\eqref{fact_def}, we have
\begin{align}
\nu_{\mathbf{s}} &= \left\langle \left[ n_1^{s_1}-A_1n_1^{s_1-1}+B_1n_1^{s_1-2}+\ldots \right]\cdots \left[ n_d^{s_d}-A_dn_d^{s_d-1}+B_dn_d^{s_d-2}+\ldots \right]\right\rangle \nonumber \\
&= \mu_{\mathbf{s}} - \sum\limits_{i=1}^d A_i\mu_{\mathbf{s}}^{(i)} + \frac{1}{2}\sum\limits_{\substack{i,j=1\\i\neq j}}^d A_iA_j\mu_{\mathbf{s}}^{(i,j)} + 
\sum\limits_{i=1}^d B_i\mu_{\mathbf{s}}^{(i,i)} + \ldots ,
\label{mu_nu}
\end{align}
where $A_i=s_i(s_i-1)/2$, and $\mu_{\mathbf{s}}^{(i)}$ is $\mu_{\mathbf{s}}$ with $s_i$ reduced by 1; similarly $\mu_{\mathbf{s}}^{(i,j)}$ is $\mu_{\mathbf{s}}$ with both $s_i$ and $s_j$ reduced by 1.

Applying \eqref{mu_nu} to that case of $\bm{s}$, but with $s_i$ reduced by 1, $s_i$ and $s_j$ are reduced by 1,..., and also noting from \eqref{fact_result} that $\nu_{\mathbf{s}}$ is of order $N^S$, we see that $\mu_{\mathbf{s}}$ is of order $N^S$, $\mu_{\mathbf{s}}^{(i)}$ is of order $N^{S-1}$ and $\mu_{\mathbf{s}}^{(i,j)}$ is of order $N^{S-2}$. Therefore 
\begin{equation}
\mu_{\mathbf{s}} = \nu_{\mathbf{s}}+\sum\limits_{i=1}^d A_i\mu_{\mathbf{s}}^{(i)}+\mathcal{O}\left(N^{S-2}\right) = \nu_{\mathbf{s}}+\sum\limits_{i=1}^d A_i\nu_{\mathbf{s}}^{(i)}+\mathcal{O}\left(N^{S-2}\right),
\end{equation}
and so
\begin{equation}
\mu_{\mathbf{s}} = N^S p_1^{s_1}\cdots p_d^{s_d}-\frac{1}{2}S(S-1)N^{S-1}p_1^{s_1}\cdots p_d^{s_d}+\frac{1}{2}\sum\limits_{i=1}^ds_i(s_i-1)\,N^{S-1}p_1^{s_1} \cdots p_{i-1}^{s_{i-1}} p^{s_i - 1}_{i}p^{s_{i+1}}_{i+1} \cdots p_d^{s_d}+\mathcal{O}\left(N^{S-2}\right).
\label{mu_of_N}
\end{equation}

Equation \eqref{mu_of_N} gives the required $N$ dependence of $\mu_{\mathbf{s}}$. Substituting back into Eq.~\eqref{eqn:Jr} now gives
\begin{equation}
 J_{\mathbf{r}}(\mathbf{p}) = \prod^{d}_{i=1}  p_i^{r_i}\,\sum^{r_i}_{s_i = 0} \binom{r_i}{s_i} \left(-1\right)^{r_i - s_i}\,\left[1-\frac{1}{2}S(S-1)N^{-1}+\frac{1}{2}\sum\limits_{j=1}^ds_j(s_j-1)(Np_j)^{-1}+\mathcal{O}\left(N^{-2}\right)\right].
\label{N_dep_J}
\end{equation}

The following relations can be straightforwardly verified:
\begin{align*}
 \sum\limits_{s_i=0}^{r_i}\binom{r_i}{s_i}\left(-1\right)^{r_i-s_i} &= 0\text{, for }r_i > 0 \\
\sum\limits_{s_i=0}^{r_i}\binom{r_i}{s_i}\left(-1\right)^{r_i-s_i}s_i(s_i-1) &= 0\text{, for }r_i > 2 \\
\sum\limits_{s_i=0}^{r_i}\binom{r_i}{s_i}\left(-1\right)^{r_i-s_i}s_i \sum\limits_{s_j=0}^{r_j}\binom{r_j}{s_j}\left(-1\right)^{r_j-s_j}s_j &= 0\text{, for }r_i > 1\text{ or } r_j > 1. \\
\end{align*}
Therefore for $\sum^{d}_{i=1} r_i > 2$, the first three terms in the square bracket of Eq.~\eqref{N_dep_J} give zero contribution, and so $J_{\mathbf{r}}\left(\mathbf{p} \right)$ is of order $N^{-2}$ for $\sum^{d}_{i=1} r_i > 2$.


%

\end{document}